# STOCHASTIC MODELING IN NANOSCALE BIOPHYSICS: SUBDIFFUSION WITHIN PROTEINS

By S. C. Kou[1]

*Harvard University*

Advances in nanotechnology have allowed scientists to study biological processes on an unprecedented nanoscale molecule-by-molecule basis, opening the door to addressing many important biological problems. A phenomenon observed in recent nanoscale single-molecule biophysics experiments is subdiffusion, which largely departs from the classical Brownian diffusion theory. In this paper, by incorporating fractional Gaussian noise into the generalized Langevin equation, we formulate a model to describe subdiffusion. We conduct a detailed analysis of the model, including (i) a spectral analysis of the stochastic integro-differential equations introduced in the model and (ii) a microscopic derivation of the model from a system of interacting particles. In addition to its analytical tractability and clear physical underpinning, the model is capable of explaining data collected in fluorescence studies on single protein molecules. Excellent agreement between the model prediction and the single-molecule experimental data is seen.

## 1. Introduction.

1.1. *Background*: *Nanoscale single-molecule biophysics.* It is said that the famous Richard Feynman once described seeing the images of single atoms as a "religious experience" for physicists. Recent advances in nanotechnology have turned Feynman's "religious" encounter into daily reality. In particular, scientists are now able to study biological processes on an unprecedented nanoscale molecule-by-molecule basis [cf. Moerner (2002), Nie and Zare (1997), Tamarat, Maali, Lounis and Orrit (2000), Weiss (2000), Xie and Lu (1999), Xie and Trautman (1998), Kou, Xie and Liu (2005)], thus

Received June 2007; revised October 2007.

[1]Supported in part by NSF Grant DMS-02-04674 and the NSF Career Award DMS-04-49204.

opening the door to addressing many problems that were inaccessible just a few decades ago.

Compared with traditional experiments, which involve a population of molecules, (nanoscale) single-molecule experiments offer many advantages. First, they provide experimental data with more accuracy and higher resolution because scientists can "zoom in" on individual molecules to study and measure them. Second, by following individual molecules, these single-molecule experiments can capture transient intermediates and detailed dynamics of biological processes. This type of information is rarely available from the traditional population experiments. Third, in a living cell, many important biological functions are often carried out by single molecules; thus, understanding the behavior of molecules at the individual level is of crucial importance. Many new discoveries [see, e.g., Asbury, Fehr and Block (2003), English et al. (2006), Kou et al. (2005), Lu, Xun and Xie (1998), Zhuang et al. (2002)] have emerged from the nanoscale single-molecule studies.

The technological advance also brings opportunities and challenges for stochastic modelers because the individual molecules, subject to statistical and quantum mechanics of the nanometer world, behave stochastically. Characterizing their fluctuation thus requires stochastic models [Kou (2007)]. In the current article we focus on modeling the phenomenon of subdiffusion observed in single-molecule experiments to exemplify the stochastic modeling problems in the field.

1.2. *Subdiffusion in proteins: The experimental finding.* Since Einstein's and Wiener's ground breaking works in the early 20th century, the theory of Brownian motion and diffusion processes has revolutionized not only physics, chemistry and biology, but also probability and statistics. One key characteristic of Brownian motion is that the second moment $E[x^2(t)]$, which in physics corresponds to the mean squared displacement (location) of a Brownian particle, is proportional to time $t$. In some systems [cf. Bouchaud and Georges (1990), Klafter, Shlesinger and Zumofen (1996), Sokolov, Klafter and Blumen (2002)], scientists, however, have discovered a clear departure from Brownian diffusion. The mean squared displacement $E[x^2(t)]$ there is no longer proportional to $t$, but rather $E[x^2(t)] \propto t^\alpha$, where $0 < \alpha < 1$. Because $\alpha < 1$, these movements satisfying $E[x^2(t)] \propto t^\alpha$ are defined as subdiffusion. Recent single-molecule biophysics experiments [Yang et al. (2003), Kou and Xie (2004), Min et al. (2005)] reveal that subdiffusion may be quite prevalent in biological systems.

In a 2003 *Science* paper [Yang et al. (2003)] scientists conducting single-molecule experiments on a protein–enzyme system observed this subdiffusion phenomenon. The experiment studied a protein–enzyme compound, called Fre, which is involved in the DNA synthesis of the bacterium *E. Coli*. In the reactions Fre works as a catalyst. Figure 1 shows the crystal structure of Fre, which contains two smaller structures: FAD (an electron carrier) and Tyr (an amino acid). The 3D conformation (shape) of Fre spontaneously fluctuates, and consequently, the (edge-to-edge) distance between the two substructures FAD and Tyr varies over time. It was found in the experiment that the stochastic distance fluctuation between FAD and Tyr undergoes a subdiffusion. Section 5 provides more details about the experiment.

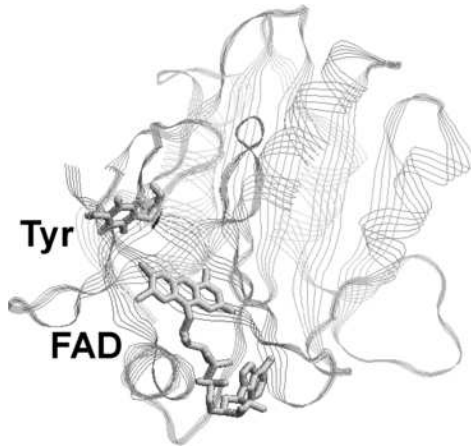

FIG. 1. *The crystal structure of Fre. The two substructures FAD and Tyr are shown.*

**1.3. *Modeling subdiffusion.*** To explain this subdiffusion phenomenon, we shall formulate a stochastic model by incorporating fractional Gaussian noise (formally the derivative of fractional Brownian motion) into a stochastic integro-differential equation framework governed by the generalized Langevin equation.

Since its introduction [Mandelbrot and Van Ness (1968)], fractional Brownian motion (fBm) has proven to be an indispensable tool for stochastic modeling [Samorodnitsky and Taqqu (1994), Taqqu (1986), Whitt (2002), Adler, Feldman and Taqqu (1998)]; its applications range from queuing systems [Glynn and Zeevi (2000), Heath, Resnick and Samorodnitsky (1997), Konstantopoulos and Lin (1996)] to finance [Heyde (1999), Mandelbrot (1997)] to internet traffic [Leland, Taqqu, Willinger and Wilson (1994), Crovella and Bestavros (1996), Mikosch, Resnick, Rootzén and Stegeman (2002)]. On the other hand, the generalized Langevin equation [Chandler (1987), Zwanzig (2001), Wang and Tokuyama (1999)], used primarily in the physics literature, has attracted less attention from probabilists and statisticians. Notably, it is the connection between fBm and the generalized Langevin equation (GLE), as we shall see, that leads to a satisfactory model to account for the experimentally observed subdiffusion within proteins.

A key requirement in the construction of biophysical models, in addition to the preference of analytical tractability, is that the model must agree with fundamental physical laws and should have a sound physical foundation. Hence, we will also discuss the model's physical basis.

From an applied probabilistic/statistical point of view, to describe subdiffusion, we study several stochastic integro-differential equations driven by fBm. We conduct a detailed analysis, in particular, a spectral analysis, of their properties (in Sections 2 and 3). We also consider the connection between our model and interacting particle systems through a microscopic derivation of the model from a system of interacting particles (in Section 4). Some of the mathematical structures of GLE and fBm have been independently considered in Kupferman (2004). In this paper, in addition to the detailed stochastic investigation, we apply the analytical results to fit the nanoscale single-molecule experimental data (in Section 5) and to show that the model successfully explains the experimental observations.

The paper is organized as follows. Section 2 concerns the basic model to describe the subdiffusive movement of a free particle. Section 3 studies subdiffusive motion under the presence of an outside potential. Section 4 investigates the physical foundation of our model. Section 5 applies the model to explain the nanoscale single-molecule experimental results, which shows close agreement between the model and the data. Section 6 concludes the paper with a discussion and raises some open problems.

## 2. Modeling subdiffusion of a free particle.

2.1. *Physical Brownian motion via the Langevin equation with white noise.* To facilitate the discussion of our model, let us first review how the law of Brownian diffusion was derived in physics because the Brownian motion used by physicists and the term Brownian motion used in probability and statistics refer to different things: Physicists' Brownian motion corresponds to the integral of the Ornstein–Uhlenbeck process, as we shall see shortly, whereas statisticians and probabilists' Brownian motion refers to the Wiener process, although both share the characteristic of $E[x^2(t)] \propto t$ for large $t$.

Suppose we have a Brownian particle with mass $m$ suspended in water. The physical description of the particle's free motion starts from the Langevin equation [Risken (1989), Van Kampen (2001), Karlin and Taylor (1981)]

$$(2.1) \qquad m\frac{dv(t)}{dt} = -\zeta v(t) + F(t),$$

where $v(t)$ is the velocity of the particle at time $t$, and $dv(t)/dt$ is the acceleration of the particle. On the right-hand side, $\zeta$ is the friction constant, reflecting the fact that the resistance the particle receives is proportional to its velocity, and $F(t)$ is the white noise, formally the derivative of the Wiener process.

The Langevin equation has an important physical constraint that links the friction constant $\zeta$ with the noise level, because both the movement of the particle and the friction originate from one source—the collision between the particle and surrounding water molecules. Borrowing physicists' notation $\delta(\cdot)$ of Dirac's delta function, this link can be expressed as

$$(2.2) \qquad E[F(t)F(s)] = 2\zeta k_B T \cdot \delta(t-s),$$

where $k_B$ is the Boltzmann constant, and $T$ is the underlying temperature. This proportional relationship between the noise level and the friction constant is a consequence of the *fluctuation–dissipation* theorem in statistical mechanics [Chandler (1987), Hill (1986)]. In the more familiar probability notation, equations (2.1) and (2.2) translate to

$$(2.3) \qquad m\,dv(t) = -\zeta v(t)\,dt + \sqrt{2\zeta k_B T}\,dB(t),$$

where $B(t)$ is the Wiener process, and the formal association of "$F(t) = \sqrt{2\zeta k_B T}\,dB(t)/dt$" is recognized.

The stationary solution of equation (2.3) is the Ornstein–Uhlenbeck process [Karlin and Taylor (1981)], which is Gaussian with mean function $E[v(t)] = 0$ and covariance function $E[v(t)v(s)] = \frac{k_B T}{m}\exp(-\frac{\zeta}{m}|t-s|)$. It follows that, for the displacement, $x(t) = \int_0^t v(s)\,ds$, which is the actual observed motion, the variance is

$$
\begin{aligned}
\mathrm{Var}[x(t)] = E[x^2(t)] &= \int_0^t \int_0^t E[v(s)v(u)]\,du\,ds \\
(2.4) \qquad &= 2\frac{k_B T}{\zeta}t - 2\frac{k_B T m}{\zeta^2}(1 - e^{-(\zeta/m)t}) \\
&\sim 2\frac{k_B T}{\zeta}t, \qquad \text{for large } t.
\end{aligned}
$$

The last line is known (in physics) as Einstein's Brownian diffusion law (where $2k_B T/\zeta$ is the Einstein diffusion constant). It is worth emphasizing that when physicists talk about the Brownian motion of a free particle, they refer to the integral of the Ornstein–Uhlenbeck process and the corresponding equation (2.4), which bears the resemblance of $\mathrm{Var}[x(t)] \propto t$ for large $t$ to the Wiener process.

2.2. *Toward subdiffusion.* The classical theory of Brownian diffusion, however, cannot explain the subdiffusion phenomenon, which, defined by $\mathrm{Var}[x(t)] \propto t^\alpha$ with $0 < \alpha < 1$ for large $t$, has been notably observed in distance fluctuation within proteins. We will explain the experimental details in Section 5, where the theoretical results and predictions of our model are compared with the experimental data. In this subsection and the next we focus on the model itself.

The starting point of our model is the *generalized Langevin equation* (GLE) [Chandler (1987), Zwanzig (2001)]

$$(2.5) \qquad m\frac{dv(t)}{dt} = -\zeta \int_{-\infty}^{t} v(u)K(t-u)\,du + G(t),$$

where, in comparison with the Langevin equation (2.1), (i) a noise $G(t)$ having memory replaces the memoryless white noise and (ii) a kernel $K$ convoluted with the velocity makes the process non-Markovian.

The reason that both $K$ and $G(t)$ appear in the equation is that any closed (equilibrium) physical system must satisfy the fluctuation–dissipation theorem, which requires the memory kernel $K(t)$ and the fluctuating noise to be linked by

$$(2.6) \qquad E[G(t)G(s)] = k_B T\zeta \cdot K(t-s).$$

In an intuitive sense this relationship arises because both the friction and the motion of the particle originate from the collision between the particle and its surrounding media. Equations (2.5) and (2.6) contain their Langevin counterpart (2.1) and (2.2) as a special case, in which the kernel $K$ is the delta function.

It is important to note that relationship (2.6) also rules out models like $m\,dv(t)/dt = -\zeta v(t) + G(t)$, with $G(t)$ an arbitrary noise to describe biophysical processes because this kind of model violates the fluctuation–dissipation theorem, and thus cannot correspond to any (equilibrium) physical process. In other words, to have a physically meaningful description of biophysical processes, the subdiffusion process, in particular, the convolution term $\int_{-\infty}^{t} v(u)K(t-u)\,du$ must be present in equation (2.5).

Under the framework of GLE, the key question is as follows: Is there a combination of kernel function and noise structure that can lead to subdiffusion? To answer this question, we note that the white noise is mathematically interpreted as the formal derivative of the Wiener process. It is well known that the Wiener process is the unique process characterized by (i) being Gaussian, (ii) having independent increments, (iii) having stationary increments, and (iv) being self-similar. These properties carry their physical meanings: the independent increments of the Wiener process make the white noise independent across time; the stationary increments of the Wiener process mean that the white noise is time translation invariant; the self-similarity means that the white noise is invariant to time scale change. To generalize the white noise, we want to maintain as many nice properties as possible and at the same time introduce memory. This leads us to consider processes with the following three properties: (i) Gaussian, (ii) stationary increments, and (iii) self-similar. The only class of processes that embodies all three properties is the fBm process, $B_H(t)$ [Embrechts and Maejima (2002), Samorodnitsky and Taqqu (1994)], which is Gaussian with mean $E[B_H(t)] = 0$ and covariance function $E[B_H(t)B_H(s)] = \frac{1}{2}(|t|^{2H} + |s|^{2H} - |t-s|^{2H})$. $H$, between 0 and 1, is the Hurst parameter; $B_H(t)$ reduces to the Wiener process when $H = 1/2$.

2.3. *Subdiffusion via the generalized Langevin equation with fractional Gaussian noise.* Taking $G(t)$ in (2.5) to be the (formal) derivative of fBm, which is referred to as the fractional Gaussian noise (fGn), $F_H(t) = \sqrt{2\zeta k_B T} \, dB_H(t)/dt$, we have the following model:

**The model for subdiffusion:**

$$(2.7) \qquad m\frac{dv(t)}{dt} = -\zeta \int_{-\infty}^{t} v(u) K_H(t-u) \, du + F_H(t),$$

where the kernel $K_H(t)$, according to equation (2.6), is (formally) given by

$$
\begin{aligned}
(2.8) \quad K_H(t) &= E[F_H(0)F_H(t)]/(k_B T \zeta) \\
&= 2\lim_{h\downarrow 0} E\left[\frac{B_H(h)}{h} \frac{B_H(t+h) - B_H(t)}{h}\right] \\
&= 2\lim_{h\downarrow 0} \frac{1}{2h^2}(|t+h|^{2H} + |t-h|^{2H} - 2|t|^{2H}) \\
&= 2H(2H-1)|t|^{2H-2}, \qquad \text{for } t \neq 0.
\end{aligned}
$$

In the more familiar probability notation equation (2.7) can be written as follows:

**The model for subdiffusion:**

$$(2.9) \quad m\,dv(t) = -\zeta\left(\int_{-\infty}^{t} v(u) K_H(t-u)\,du\right)dt + \sqrt{2\zeta k_B T}\,dB_H(t),$$

where, as in (2.8),

$$(2.10) \qquad K_H(t) = 2H(2H-1)|t|^{2H-2}, \qquad \text{for } t \neq 0.$$

One question about this model equation (2.9) arises immediately from a probabilistic standpoint: what is the interpretation of the integral with respect to fBm?

Stochastic integrals driven by fBm have been an active research area in recent years. The constructions of the stochastic integral include restricting the integrand to specific classes of functions as in Dai and Heyde (1996), Gripenberg and Norros (1996), Lin (1995), Shiryaev (1998), using pathwise integration for the case of $1/2 < H < 1$ as in Mikosch and Norvaisa (2000), applying Malliavin calculus as in Alòs, Mazet and Nualart (2000),

Duncan, Hu and Pasik-Duncan (2000), Nualart (2006), and using regularization as in Rogers (1997), Carmona and Coutin (2000). For reviews, see, for example, Duncan, Hu and Pasik-Duncan (2000), Pipiras and Taqqu (2000), Pipiras and Taqqu (2001) and Embrechts and Maejima (2002).

As we shall see shortly, the case of $1/2 < H < 1$ is particularly relevant to our description of subdiffusion here. In this case, pathwise integration appears most natural, and hence, we interpret our model equation (2.9) with $H > 1/2$ in the pathwise Riemann–Stieltjes sense [Mikosch and Norvaisa (2000)]. This pathwise integration allows us to treat integrals with respect to $dB_H(t)$ as if they were classical integrals, which simplifies our calculation.

The presence of the convolution term and the $dB_H(t)$ term makes equation (2.9) non-Markovian and nonstandard. The solution to model (2.9) is given by the following theorem, whose proof is deferred to the Appendix.

THEOREM 2.1.  Let $\tilde{K}_H(\omega)$ and $\tilde{K}_H^+(\omega)$ denote the Fourier transforms of the kernel $K_H(t)$ on the entire and positive real lines, respectively:

$$(2.11) \quad \tilde{K}_H(\omega) = \int_{-\infty}^{\infty} e^{it\omega} K_H(t)\, dt = 2\Gamma(2H+1)\sin(H\pi)|\omega|^{1-2H},$$

$$\begin{aligned}(2.12) \quad \tilde{K}_H^+(\omega) &= \int_0^{\infty} e^{it\omega} K_H(t)\, dt \\ &= \Gamma(2H+1)|\omega|^{1-2H}[\sin(H\pi) - i\cos(H\pi)\mathrm{sign}(\omega)],\end{aligned}$$

where $\mathrm{sign}(\omega)$ is the sign function, which is 1 if $\omega > 0$ and $-1$ if $\omega < 0$. Then under the pathwise interpretation of $dB_H(t)$ for $1/2 < H < 1$, the solution to model (2.9) is

$$v(t) = \sqrt{2\zeta k_B T} \int_{-\infty}^{\infty} r(t-u)\, dB_H(u),$$

where the deterministic function $r(t)$ is given by

$$r(t) = \frac{1}{2\pi} \int_{-\infty}^{\infty} \frac{1}{\zeta \tilde{K}_H^+(\omega) - im\omega} e^{-it\omega}\, d\omega.$$

Furthermore, the solution $v(t)$ is a stationary Gaussian process with mean function $E[v(t)] = 0$ and covariance function

$$C_v(t) = E[v(0)v(t)] = \frac{1}{2\pi} \int_{-\infty}^{\infty} e^{-it\omega} \tilde{C}_v(\omega)\, d\omega,$$

where the Fourier transform $\tilde{C}_v(\omega)$ of $C_v(t)$ is given by

$$\tilde{C}_v(\omega) = \int_{-\infty}^{\infty} e^{it\omega} C_v(t)\, dt = k_B T \zeta \tilde{K}_H(\omega)/|\zeta \tilde{K}_H^+(\omega) - im\omega|^2.$$

REMARK 1. When $H \to 1/2$, we have $\tilde{C}_v(\omega) \to 2k_BT\zeta/(\zeta^2 + m^2\omega^2)$ and $E[v(0)v(t)] = C_v(t) \to \frac{k_BT}{m}\exp(-\frac{\zeta}{m}|t|)$, which recovers the Ornstein–Uhlenbeck result from classical Brownian diffusion.

REMARK 2. A careful reader might note that equations (2.11) and (2.12) involve Fourier transforms of power functions, which are not integrable, so a natural question here is their meaning. In general, the Fourier transform $\tilde{f}$ of a nonintegrable function $f$ is defined as $\tilde{f}(\omega) = \lim_{\alpha \to \infty} \int_{-\infty}^{\infty} f(t)\exp(-|t|/\alpha) \times \exp(it\omega)\,dt$, that is, the limit of exponential damping. It is in this sense that, for example, $|t|^{-1/2}$ and $\sqrt{2\pi}|\omega|^{-1/2}$ are regarded as a Fourier transform pair. See Champeney (1987) for a thorough discussion. All the Fourier and inverse Fourier transforms in this article are defined in this general sense.

The next theorem, whose proof is given in the Appendix, details how our model (2.9) explains subdiffusion.

THEOREM 2.2. Under model (2.9), let $x(t) = \int_0^t v(s)\,ds$ be the displacement. Then for $1/2 < H < 1$, the mean squared displacement

$$\text{(2.13)} \quad \text{Var}[x(t)] = E[x(t)^2] \sim \frac{k_BT}{\zeta}\frac{\sin(2H\pi)}{\pi H(1-2H)(2-2H)}t^{2-2H}$$

$$\propto t^{2-2H}, \qquad \text{for large } t.$$

In other words, our model with $1/2 < H < 1$ leads to subdiffusion.

REMARK 3. When $H \to 1/2$, the right-hand side of (2.14) converges to $(2k_BT/\zeta)t$, and we recover the Einstein Brownian diffusion law.

The above theorem says that, for a particle, if its velocity process $v(t)$ is governed by model (2.9) with $H > 1/2$, then the second moment $E[x^2(t)]$ of the displacement $x(t) = \int_0^t v(s)\,ds$ satisfies $E[x^2(t)] \propto t^{2-2H}$ for large $t$, which is exactly the characteristic of subdiffusion. Therefore, our model (2.9) with Hurst parameter $H > 1/2$ leads to an explanation of how subdiffusion could arise in biophysical systems—as long as the fractional Gaussian noise drives the underlying physical fluctuation, the system will be subdiffusive.

## 3. Modeling subdiffusion under external potential.
The model (2.9) explains subdiffusion of a free particle, that is, the motion of a particle without the influence of an outside force (or potential). If there is an external potential $U(x)$ (e.g., a magnetic field), which is a function of the displacement $x(t)$, the model has to be modified.

In the case of white noise and the Langevin equation (2.1), which corresponds to classical Brownian diffusion, one should add the term $-U'(x(t))$ to the right-hand side of the equation to account for the external potential [Risken (1989), Van Kampen (2001), Karlin and Taylor (1981)]:

$$m\frac{dv(t)}{dt} = -\zeta v(t) - U'(x(t)) + F(t), \qquad x(t) = \int_0^t v(s)\,ds.$$

For GLE, similarly, the term $-U'(x(t))$ also needs to be added to the right-hand side of the equation to account for the external potential [Chandler (1987), Zwanzig (2001)]. Therefore, to describe the movement of a subdiffusive particle under the presence of an external potential $U(x)$, we have the following model.

**The model for subdiffusion under a general potential $U(x)$:**

$$m\frac{dv(t)}{dt} = -\zeta \int_{-\infty}^t v(u)K_H(t-u)\,du - U'(x(t)) + F_H(t),$$

$$x(t) = \int_0^t v(s)\,ds,$$

which is the companion of (2.7).

The harmonic potential $U(x) = \frac{1}{2}m\psi x^2$, where $m$ is the mass of the particle and the constant $\psi$ reflects the potential's strength, is of particular importance because when the movement is confined to a short range, such as the movement within a protein as we shall see in Section 5, the underlying potential function can often be adequately approximated by a harmonic one. Under such a harmonic potential, the model is as follows.

**The model for subdiffusion under a harmonic potential:**

(3.1)
$$m\frac{dv(t)}{dt} = -\zeta \int_{-\infty}^t v(u)K_H(t-u)\,du - m\psi x(t) + F_H(t),$$

$$x(t) = \int_0^t v(s)\,ds.$$

An equivalent expression in the more familiar probability notation is

$$dx(t) = v(t)\,dt,$$

(3.2)
$$m\,dv(t) = -\zeta\left(\int_{-\infty}^t v(u)K_H(t-u)\,du\right)dt$$

$$- m\psi x(t)\,dt + \sqrt{2\zeta k_B T}\,dB_H(t).$$

REMARK 4. Compared with the model (2.9) for the movement of a free particle, the above model (3.2) has two distinctive features. First, the presence of the external potential changes the model from a single equation to a set of two coupled equations, and correspondingly, the solution to model (3.2) is a two-dimensional process. Second, for a free particle, since there is no external potential to bound its movement, the displacement process $x(t)$ cannot be stationary [which is manifested in Theorem 2.2, where the variance $\mathrm{Var}(x(t)) \propto t^{2-2H} \to \infty$, as $t \to \infty$]. On the other hand, for model (3.2), since the harmonic potential always pulls the particle back to the origin [because of the term $-m\psi x(t)$], the displacement process $x(t)$ can be stationary, and thus, we can talk about the stationary mean and variance of $x(t)$, as we shall see next.

The next theorem, whose proof is deferred to the Appendix, gives the solution to model (3.2), which describes the subdiffusive motion under a harmonic potential.

THEOREM 3.1. *Under the pathwise interpretation of $dB_H(t)$ for $1/2 < H < 1$, the solution to equation (3.2) is*

$$x(t) = \sqrt{2\zeta k_B T} \int_{-\infty}^{\infty} \rho(t-u)\, dB_H(u),$$

$$v(t) = \sqrt{2\zeta k_B T} \int_{-\infty}^{\infty} \rho'(t-u)\, dB_H(u),$$

*where the deterministic function $\rho(t)$ is defined as*

$$\rho(t) = \frac{1}{2\pi} \int_{-\infty}^{\infty} e^{-it\omega} \frac{1}{m\psi - m\omega^2 - i\omega\zeta \tilde{K}_H^+(\omega)}\, d\omega.$$

*Furthermore, the solution $(x(t), v(t))$ is a stationary bivariate Gaussian process with mean function $E[x(t)] = E[v(t)] = 0$ and covariance functions given by*

$$E[x(s)x(s+t)] = E[x(0)x(t)]$$
$$= \frac{k_B T \zeta}{2\pi} \int_{-\infty}^{\infty} e^{-it\omega} \frac{\tilde{K}_H(\omega)}{|m\psi - m\omega^2 - i\omega\zeta \tilde{K}_H^+(\omega)|^2}\, d\omega,$$

$$E[x(s)v(s+t)] = E[x(0)v(t)]$$
$$= \frac{k_B T \zeta}{2\pi} \int_{-\infty}^{\infty} e^{-it\omega} \frac{i\omega\tilde{K}_H(\omega)}{|m\psi - m\omega^2 - i\omega\zeta \tilde{K}_H^+(\omega)|^2}\, d\omega,$$

$$E[v(s)x(s+t)] = E[v(0)x(t)]$$
$$= \frac{k_B T \zeta}{2\pi} \int_{-\infty}^{\infty} e^{-it\omega} \frac{i\omega\tilde{K}_H(\omega)}{|m\psi - m\omega^2 - i\omega\zeta \tilde{K}_H^+(\omega)|^2}\, d\omega,$$

$$E[v(s)v(s+t)] = E[v(0)v(t)]$$

$$= \frac{k_B T \zeta}{2\pi} \int_{-\infty}^{\infty} e^{-it\omega} \frac{\omega^2 \tilde{K}_H(\omega)}{|m\psi - m\omega^2 - i\omega\zeta \tilde{K}_H^+(\omega)|^2} \, d\omega,$$

where the expressions of $\tilde{K}_H(\omega)$ and $\tilde{K}_H^+(\omega)$ are given by (2.11) and (2.12), respectively.

REMARK 5. It is straightforward to verify that in the limiting case of $\psi \to 0$, the result of $E[v(0)v(t)]$ in the above theorem converges to the expression of $C_v(t)$ in Theorem 2.1. This says that as the harmonic potential becomes weaker and weaker, the particle will behave more and more like a free particle, and in the limit the movement reduces to that of a free particle.

One special case of model (3.1) is when the acceleration term $m\, dv(t)/dt$ is negligible. This corresponds to the so-called overdamped condition in physics [Van Kampen (2001)], where the friction in the system is very large, causing the acceleration of the particle to be negligible. In the overdamped scenario, the acceleration term drops out, and the model changes to the following.

**The model for subdiffusion under a harmonic potential and the overdamped condition:**

$$(3.3) \quad m\psi x(t) = -\zeta \int_{-\infty}^{t} v(u) K_H(t-u) \, du + F_H(t), \qquad x(t) = \int_0^t v(s) \, ds.$$

It can be rewritten in the more familiar probability notation as

$$(3.4) \quad \begin{aligned} dx(t) &= v(t)\, dt, \\ m\psi x(t)\, dt &= -\zeta \left( \int_{-\infty}^{t} v(u) K_H(t-u) \, du \right) dt + \sqrt{2\zeta k_B T} \, dB_H(t). \end{aligned}$$

The next theorem (with proof given in the Appendix) solves equation (3.4). We will use the solution to explain the experimental data in Section 5 because in biological systems, such as within a protein, the friction is usually large and the overdamped condition usually holds.

THEOREM 3.2. *Under the pathwise interpretation of the stochastic integral for $1/2 < H < 1$, the solution to equation (3.4) is a stationary Gaussian process with mean $E[x(t)] = 0$ and covariance function*

$$(3.5) \qquad \sigma_x(t) = E[x(0)x(t)] = \frac{k_B T}{m\psi} E_{2-2H}(-(t/\tau)^{2-2H}),$$

*where the constant*

$$(3.6) \qquad \tau = \left( \frac{\zeta \Gamma(2H+1)}{m\psi} \right)^{1/(2-2H)},$$

*and $E_\alpha(z)$ is the Mittag–Leffler function [see Erdélyi et al. (1953), Chapter 18] defined by*

$$E_\alpha(z) = \sum_{k=0}^{\infty} z^k / \Gamma(\alpha k + 1).$$

REMARK 6.   The Mittag–Leffler function generalizes the exponential function in a natural way. When $H \to 1/2$, the Mittag–Leffler function in (3.5) reduces to the exponential function, and $\sigma_x(t) = (k_B T / m\psi) \exp(-(m\psi/\zeta)t)$, recovering the classical Brownian diffusion result.

**4. Physical basis of the model.**   We shall apply the results in the previous two sections to explain the nanoscale subdiffusive motion observed within proteins. But before doing so, we will study in this section the physical foundation of the model, since a key requirement for biophysical models is that, in addition to satisfying fundamental physical laws, they must have a sound physical basis.

4.1. *The thermal dynamic requirement for a free particle.*   The law of thermal dynamics [Chandler (1987), Hill (1986), Reif (1965)] requires that, for a free particle, the (equilibrium) stationary variance of its velocity should be $k_B T / m$, where $m$ is the mass of the particle. The next theorem (whose proof is deferred to the Appendix) verifies that indeed our model (2.9) for the free particle satisfies this thermal dynamic requirement.

THEOREM 4.1.   *Under model* (2.9), *the stationary variance of the velocity* $\mathrm{Var}[v(0)]$ *satisfies*

$$\mathrm{Var}[v(0)] = \frac{k_B T}{m} \qquad \text{for all } 1/2 < H < 1.$$

4.2. *The thermal dynamic requirement for the movement of a particle under harmonic potential.*   For particles moving under a harmonic potential $U(x) = \frac{1}{2} m\psi x^2$, the law of thermal dynamics asserts that the equilibrium (stationary) variance of the displacement should be $\frac{k_B T}{m\psi}$. The next theorem (whose proof is given in the Appendix) confirms that our model (3.2) for harmonic potential indeed satisfies the thermal dynamic requirement.

THEOREM 4.2.   *Under model* (3.2), *the stationary variance of the displacement* $\mathrm{Var}[x(0)]$ *satisfies the thermal dynamic requirement of*

$$\mathrm{Var}[x(0)] = \frac{k_B T}{m\psi} \qquad \text{for all } 1/2 < H < 1.$$

For model (3.4), which describes subdiffusion under harmonic potential and the overdamped condition, the results of Theorem 3.2 imply

$$(4.1) \quad \sigma_x(0) = \text{Var}[x(0)] = \frac{k_B T}{m\psi} E_{2-2H}(0) = \frac{k_B T}{m\psi} \qquad \text{for all } 1/2 < H < 1.$$

We thus have the following:

THEOREM 4.3. *Under model* (3.4), *the stationary variance of the displacement* $\text{Var}[x(0)]$ *satisfies the thermal dynamic requirement of*

$$\text{Var}[x(0)] = \frac{k_B T}{m\psi} \qquad \text{for all } 1/2 < H < 1.$$

4.3. *Deriving the model from a system of interacting particles.* In this subsection we will demonstrate that the model in Section 3 can be derived from the physical microscopic interaction between the particle under study and its surrounding media; in particular, the model can be derived from a Hamiltonian system consisting of the particle and its surroundings. For more general discussion about the Hamiltonian and GLE, see Zwanzig (2001).

We start the derivation from the Hamiltonian [Corben and Stehle (1995)] of the particle

$$H_s = \frac{p^2}{2m} + \frac{1}{2} m\psi x^2,$$

where $p = mv$ is the momentum, and $x$ is the displacement. Here the Hamiltonian, which is essentially the total energy of the particle, consists of the kinetic energy $p^2/(2m) = mv^2/2$ and the potential energy, which is $m\psi x^2/2$ under the harmonic case. The surrounding media, consisting of $N$ small molecules, has its own Hamiltonian (total energy)

$$(4.2) \quad H_B = \sum_{j=1}^{N} \left( \frac{p_j^2}{2m_b} + \frac{1}{2} m_b \omega_j^2 \left( q_j - \frac{\gamma_j}{\omega_j^2} x \right)^2 \right),$$

where $N$, on the order of $10^{23}$, is the total number of molecules in the media, $m_b$ is the (common) mass of an individual molecule in the media, $p_j$ and $q_j$ are respectively the momentum and location of the $j$th molecule, $\omega_j$ is the oscillation frequency of the $j$th molecule (as each individual molecule oscillates in the media), and the term $(q_j - \gamma_j x/\omega_j^2)$ captures the interaction between the particle of interest and the individual molecules in the media, where $\gamma_j$ is the interacting strength between the particle and the $j$th molecule.

The total Hamiltonian (energy) of the entire system (i.e., the particle plus the media) is thus $H_s + H_B$. Once the total Hamiltonian is given, the classical theory of mechanics [Corben and Stehle (1995)] states that the motion of the particle, as well as that of the individual molecules, is given by

$$(4.3) \qquad \frac{dx}{dt} = \frac{\partial(H_s + H_B)}{\partial p}, \qquad \frac{dp}{dt} = -\frac{\partial(H_s + H_B)}{\partial x},$$

$$(4.4) \qquad \frac{dq_j}{dt} = \frac{\partial(H_s + H_B)}{\partial p_j}, \qquad \frac{dp_j}{dt} = -\frac{\partial(H_s + H_B)}{\partial q_j},$$

which is a set of coupled differential equations. The exact expressions of $H_s$ and $H_B$ reduce (4.3) and (4.4) to

$$(4.5) \qquad \frac{dx}{dt} = \frac{p}{m}, \qquad \frac{dp}{dt} = -m\psi x + \sum_{j=1}^{N} \gamma_j m_b \left( q_j - \frac{\gamma_j}{\omega_j^2} x \right),$$

$$(4.6) \qquad \frac{dq_j}{dt} = \frac{p_j}{m_b}, \qquad \frac{dp_j}{dt} = -m_b \omega_j^2 q_j + m_b \gamma_j x,$$

from which we can first express $q_j$ and $p_j$ in terms of $x(t)$ and their initial values:

$$q_j(t) = q_j(0)\cos(\omega_j t) + \frac{p_j(0)}{m_b \omega_j}\sin(\omega_j t) + \frac{\gamma_j}{\omega_j}\int_0^t x(s)\sin(\omega_j(t-s))\,ds.$$

Applying an integration by parts on it gives

$$q_j(t) - \frac{\gamma_j}{\omega_j^2} x(t) = \left[ q_j(0) - \frac{\gamma_j}{\omega_j^2} x(0) \right] \cos(\omega_j t) + \frac{p_j(0)}{m_b \omega_j}\sin(\omega_j t)$$

$$- \frac{\gamma_j}{\omega_j^2}\int_0^t \cos(\omega_j(t-s))\,dx(s).$$

Taking this expression into (4.5), we obtain

$$m\frac{d^2x}{dt^2} = -m\psi x(t) - \int_0^t J(t-s)\,dx(s) + G(t),$$

where

$$J(t) = m_b \sum_{j=1}^{N} \frac{\gamma_j^2}{\omega_j^2} \cos(\omega_j t),$$

$$G(t) = \sum_{j=1}^{N} m_b \gamma_j \left( q_j(0) - \frac{\gamma_j}{\omega_j^2} x(0) \right) \cos(\omega_j t) + \sum_{j=1}^{N} \frac{\gamma_j}{\omega_j} p_j(0)\sin(\omega_j t).$$

Upon making the association

$$\zeta \propto m_b, \qquad \zeta K(t) = J(t),$$

we reach the GLE model

$$m\frac{dv(t)}{dt} = -m\psi x(t) - \zeta \int_0^t K(t-s)v(s)\,ds + G(t), \qquad x(t) = \int_0^t v(s)\,ds.$$

REMARK 7. If we let the harmonic potential $m\psi x^2/2$ become weaker and weaker, then in the limit of $\psi \to 0$, the model (2.5) of a free particle is recovered.

The randomness of $G(t)$ comes from the fact that the initial values of $\mathbf{q} = (q_1(0), q_2(0), \ldots, q_N(0))$ and $\mathbf{p} = (p_1(0), \ldots, p_N(0))$ have a thermal dynamic distribution

$$f(\mathbf{p}, \mathbf{q}) \propto \exp\left(-\frac{H_B}{k_B T}\right),$$

which leads to

$$E[p_j(0)] = E[q_j(0) - \gamma_j x(0)/\omega_j^2] = 0 \qquad \text{for all } j,$$

$$E[p_j^2(0)] = m_b k_B T, \qquad E[(q_i(0) - \gamma_j x(0)/\omega_j^2)^2] = \frac{k_B T}{m_b \omega_j^2},$$

$$E[p_j(0)(q_i(0) - \gamma_i x(0)/\omega_i^2)] = 0 \qquad \text{for all } i \text{ and } j,$$

implying

$$E[G(t)G(s)] = k_B T m_b \sum_{j=1}^N \frac{\gamma_j^2}{\omega_j^2} \cos(\omega_j(t-s)) = k_B T \zeta K(t-s),$$

which exactly recovers the fluctuation–dissipation relationship (2.6).

**The underpinning of a fractional Gaussian memory kernel.** Because $N$, the total number of molecules in the media, is large, the memory kernel $K(t) \propto \frac{1}{N}\sum_{j=1}^N \gamma_j^2 \cos(\omega_j t)/\omega_j^2$ is essentially given by

$$K(t) \propto E\left[\frac{\gamma^2}{\omega^2}\cos(\omega t)\right] = \int_0^\infty \frac{E[\gamma^2|\omega]}{\omega^2}\cos(\omega t)g(\omega)\,d\omega,$$

where $g(\cdot)$ is the probability density function of the molecules' oscillation frequencies. If $\gamma$ and $g(\omega)$ are such that

$$(4.7) \qquad \frac{E[\gamma^2|\omega]}{\omega^2}g(\omega) \propto \omega^{1-2H},$$

then

$$K(t) \propto |t|^{2H-2} \propto K_H(t),$$

giving rise to the memory kernel (2.10) of the fractional Gaussian noise. Many scenarios can lead to equation (4.7). For example, if the interacting strength $\gamma$ is a (deterministic) power function of $\omega$: $\gamma^2 \approx \omega^{3-2H}$, and the distribution $g$ is roughly uniform over the spectrum, then (4.7) would hold approximately. Another example is $\gamma$ and $\omega$ being independent and $g$ having an (approximate) power tail $g(\omega) \sim \omega^{3-2H} \exp(-\alpha\omega)$ with a very small $\alpha$; then (4.7) would also hold approximately.

**5. From theory to experiments.** A recent single-molecule experiment [Yang et al. (2003)] studied a protein–enzyme compound Fre, which is involved in the DNA synthesis of *E. Coli.* As shown in Figure 1, Fre contains two subunits: FAD and Tyr. Because the 3D conformation of Fre spontaneously fluctuates over time, the (edge-to-edge) distance between FAD and Tyr varies. This distance is one dimensional; its fluctuation provides information about the conformational dynamics of Fre. To experimentally probe this one-dimensional distance fluctuation, Fre is placed under a laser beam. The laser excites FAD to be fluorescent. By recording the fluorescence lifetime of FAD, one can trace the distance between FAD and Tyr, because at any time $t$ the fluorescence lifetime $\lambda(t)$ of FAD is a function of the one-dimensional distance [see Gray and Winkler (1996), Moser et al. (1992)]

$$(5.1) \qquad \lambda(t) = k_0 e^{\beta(x_{eq}+x(t))},$$

where $k_0$ and $\beta$ are known constants [Moser et al. (1992)], $x_{eq}$ is the mean distance, and $x(t)$ with mean 0 is the distance fluctuation at time $t$.

To model $x(t)$, we first note that the external potential experienced by the fluctuating subunits is well approximated by a harmonic one, $U(x) = m\psi x^2/2$, because the movement is confined within the short range of Fre. We shall see in Section 5.3 that this close approximation is well tested in the experiment.

5.1. *Testing the autocorrelation structure of the model.* With the harmonic potential, people used to model $x(t)$ as a Brownian diffusion process $m\frac{d}{dt}v(t) = -\zeta v(t) - m\psi x(t) + F(t), x(t) = \int_0^t v(s)\,ds$, or by its overdamped version $m\psi x(t) = -\zeta v(t) + F(t), x(t) = \int_0^t v(s)\,ds$, where $F(t)$ is the white noise.

The nanoscale single-molecule experimental data of $\lambda(t)$, unlike the traditional population experiments, provides the means to test the model. One can calculate the empirical autocorrelation function of $\lambda(t)$ from the experimental data and compare it with the theoretical autocorrelation function from the model. The autocorrelation function is used as the test statistic because the experimentally recorded fluorescence lifetime is actually the true $\lambda(t)$ plus background and equipment noise. Doing an autocorrelation effectively removes the noise (since the noise is uncorrelated). For a stationary

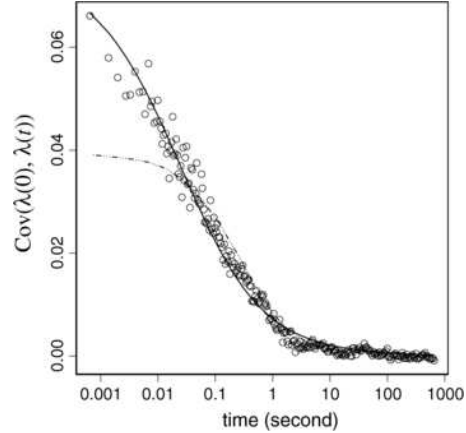

Fig. 2. *Autocorrelation function of the fluorescence lifetime* $\lambda(t)$. *The open circles represent the empirical autocorrelation calculated from the experimental data. The dashed line is the best fit from the classical Brownian diffusion model. The solid line is the fit [$H = 0.74$, $\zeta/(m\psi) = 0.40$, $\beta^2 k_B T/(m\psi) = 0.81$] from our model (3.4), agreeing well with the data.*

Gaussian process $x(t)$, it is straightforward to calculate the autocorrelation function of $\lambda(t)$ from equation (5.1),

$$(5.2) \qquad \mathrm{cov}(\lambda(0), \lambda(t)) = k_0^2 e^{2\beta x_{eq} + \beta^2 C_x(0)} (e^{\beta^2 C_x(t)} - 1),$$

where $C_x(t) = \mathrm{cov}(x(0), x(t))$. Figure 2 shows the empirical autocorrelation function (the open circles) compared with the best (least-square) fitting from the Brownian diffusion model (the dashed curve). A clear discrepancy is seen.

The solid line in Figure 2 is the result from modeling $x(t)$ by the subdiffusive process (3.4) under the harmonic potential. The curve is fitted by using the Hurst parameter $H = 0.74$, expression (5.2) and the result of Theorem 3.2 [with $\sigma_x(t)$ replacing $C_x(t)$ in (5.2)]. A very close agreement with the experimental autocorrelation function is seen. Here the overdamped model (3.4) is applied to explain the data, since the movement within a protein is subject to the overdamped regime.

5.2. *Testing higher-order correlation functions.* To check our model, we make predictions about the distance fluctuation and test whether these predictions can be confirmed by the experiments. The first set of predictions involves higher-order autocorrelation functions because they are very sensitive to distinguishing models [Mukamel (1995)]. With the values of the fitting parameters fixed to those in Figure 2, we compute from the model the predicted three-step and four-step autocorrelation functions $E[\Delta\lambda(0)\Delta\lambda(t_1)\Delta\lambda(t_1 + t_2)]$ and $E[\Delta\lambda(0)\Delta\lambda(t_1)\Delta\lambda(t_1 + t_2)\Delta\lambda(t_1 + t_2 + t_3)]$, where $\Delta\lambda(t) = \lambda(t) -$ $E[\lambda(t)]$, and compare them with their experimental counterparts. The exact expressions for the three-step and four-step autocorrelation functions are given in the Appendix.

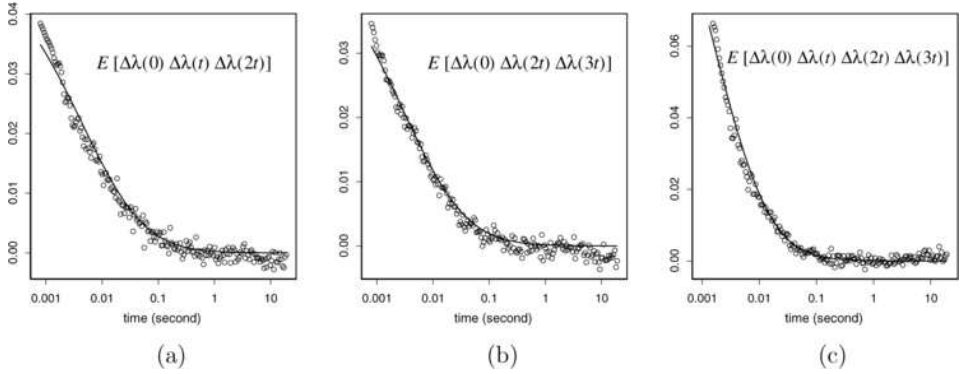

FIG. 3. *Higher-order autocorrelation functions of $\lambda(t)$.* (a), (b) *and* (c): *The experimentally obtained autocorrelation functions $E[\Delta\lambda(0)\Delta\lambda(t)\Delta\lambda(2t)]$, $E[\Delta\lambda(0)\Delta\lambda(2t)\Delta\lambda(3t)]$ and $E[\Delta\lambda(0)\Delta\lambda(t)\Delta\lambda(2t)\Delta\lambda(3t)]$ overlaid with the model predictions for various $t$. The theoretical curves from the model (3.4) are calculated using the same parameter values as in Figure 2 [$H = 0.74$, $\zeta/(m\psi) = 0.40$, and $\beta^2 k_B T/(m\psi) = 0.81$].*

Figure 3(a) shows the evenly spaced three-step autocorrelation function $E[\Delta\lambda(0)\Delta\lambda(t)\Delta\lambda(2t)]$ as a function of time $t$; Figure 3(b) shows the unevenly spaced three-step autocorrelation function $E[\Delta\lambda(0)\Delta\lambda(2t)\Delta\lambda(3t)]$ as a function of time $t$; Figure 3(c) shows the evenly spaced four-step autocorrelation function $E[\Delta\lambda(0)\Delta\lambda(t)\Delta\lambda(2t)\Delta\lambda(3t)]$ as a function of $t$. The theoretical curves (the solid lines) in Figure 3 are calculated from model (3.4) using the parameter values obtained from the fitting in Figure 2. In all cases, the close agreement between the theoretical curves (the solid lines) and the experimental values (the open circles) is seen.

The second prediction from the model is time-symmetry. For any $t_1$ and $t_2$, the model predicts $E[\Delta\lambda(0)\Delta\lambda(t_1)\Delta\lambda(t_1 + t_2)] = E[\Delta\lambda(0)\Delta\lambda(t_2)\Delta\lambda(t_1 + t_2)]$, which can be straightforwardly seen from the formulas in the Appendix. It says that if our model is true, then one can swap the order of the time lags without changing the correlation value. This can be tested by taking $t_1 = t$, $t_2 = 2t$ and plotting the experimentally obtained three-step correlation $E[\Delta\lambda(0)\Delta\lambda(t)\Delta\lambda(3t)]$ against $E[\Delta\lambda(0)\Delta\lambda(2t)\Delta\lambda(3t)]$ for various $t$. A 45° line is predicted by the model. The experimental plot in Figure 4 indeed confirms the prediction.

5.3. *Testing the harmonic potential.* As a final check of our model, we ask if the two important model assumptions of harmonic potential and fractional Gaussian memory kernel (2.10) can be directly verified from the experiment.

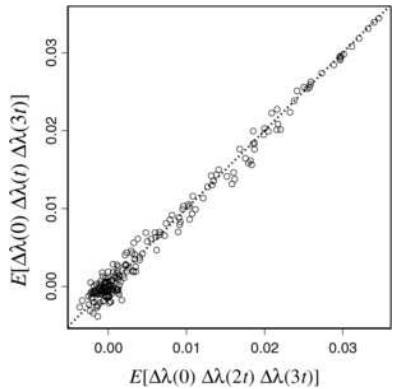

Fig. 4. *A test for time-symmetry. The experimental three-step correlations* $E[\Delta\lambda(0)\Delta\lambda(t)\Delta\lambda(3t)]$ *and* $E[\Delta\lambda(0)\Delta\lambda(2t)\Delta\lambda(3t)]$ *plotted again each other for various* $t$. *A* $45°$ *line is predicted by our model.*

Another recent single-molecule experiment [Min et al. (2005)] indeed confirmed the assumptions. This experiment studied a protein complex formed by fluorescein (FL) and monoclonal antifluorescein (anti-FL). See Figure 5. Similar to Fre, this complex contains two substructures Tyr and FL, between which the distance fluctuates over time. Using exactly the same experimental technique as in the previous Fre experiment, the distance fluctuation can be probed from the fluorescence lifetime of FL (upon placing the complex under a laser beam).

This latter experiment has identical settings as the previous one, except that it has much higher signal-to-noise ratio, and thus provides higher resolution data that allows the model assumptions to be further tested. The high resolution data on $\lambda(t)$ first enables $x(t)$ to be reconstructed from (5.1) through a local binning (kernel) average. See Figure 6(a). The empirical equilibrium (stationary) distribution $\hat{P}(x)$ of $x(t)$ is then obtained from the histogram of all the $x(t)$. According to thermal dynamics, the equilibrium distribution $P(x)$ and the potential function $U(x)$ is linked by $P(x) \propto \exp(-U(x)/(k_B T))$. A natural estimate for $U(x)$ is thus $\hat{U}(x) = -k_B T \log(\hat{P}(x))$, which is shown in Figure 6(b). A harmonic potential, the dashed line in Figure 6(b), is seen to fit $\hat{U}(x)$ very well, hence confirming the validity of our earlier assumption.

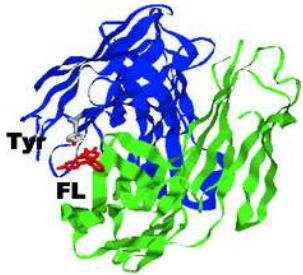

Fig. 5. *The crystal structure of the FL and anti-FL complex. The two substructures Tyr and FL are highlighted. This new experiment has identical settings as the previous Fre experiment (Figure 1), but it has much higher signal-to-noise ratio that enables experimental testing of the key assumptions of our model.*

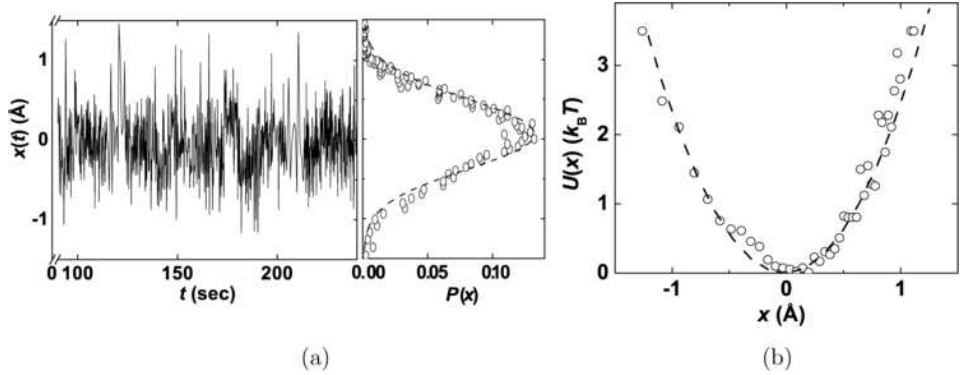

Fig. 6.  (a) *The distance fluctuation $x(t)$ reconstructed from* (5.1), *and the corresponding empirical distribution* $\hat{P}(x)$. (b) *The estimated empirical potential function* $\hat{U}(x) = -k_B T \log(\hat{P}(x))$, *the open circles, compared with the harmonic potential, the dashed line.*

5.4. *Testing the fractional Gaussian memory kernel.* To test the fractional Gaussian memory kernel (2.10), first we calculate from the reconstructed $x(t)$ the experimental autocorrelation function $\hat{C}_x(t) = \widehat{\text{cov}}(x(0), x(t))$ and compare it with the theoretical expression (3.5) from Theorem 3.2. Figure 7(a) shows the comparison, where the result from our model (the solid line) agrees well with the experimental values (the open circles).

Furthermore, from the overdamped GLE $m\psi x(t) = -\zeta \int_{-\infty}^{t} v(u)K(t - u)\,du + G(t), x(t) = \int_0^t v(s)\,ds$ with an arbitrary memory kernel $K$, we can establish a one-to-one correspondence between the Laplace transform of $C_x(t) = \text{cov}(x(0), x(t))$ and the Laplace transform of the memory kernel $K(t)$, as shown in the following theorem, which then allows us to recover $K(t)$ from the experimental $\hat{C}_x(t) = \widehat{\text{cov}}(x(0), x(t))$ and compare it with the assumed fractional Gaussian noise memory kernel.

THEOREM 5.1.  *For the overdamped GLE $m\psi x(t) = -\zeta \int_{-\infty}^{t} v(u)K(t - u)\,du + G(t), x(t) = \int_0^t v(s)\,ds$ with a general memory kernel $K(t)$, there is a one-to-one correspondence between the Laplace transform $\mathfrak{C}_x(s) =$* $\int_0^\infty e^{-st} C_x(t)\,dt$ *of* $C_x(t) = \mathrm{cov}(x(0), x(t))$ *and the Laplace transform* $\mathfrak{K}(s) = \int_0^\infty e^{-st} K(t)\,dt$ *of the memory kernel* $K(t)$:

$$(5.3)\quad \mathfrak{C}_x(s) = \frac{k_B T \zeta}{m \psi}\, \frac{\mathfrak{K}(s)}{m\psi + \zeta s \mathfrak{K}(s)}, \qquad \mathfrak{K}(s) = \frac{m\psi}{\zeta}\, \frac{m\psi \mathfrak{C}_x(s)}{k_B T - m\psi s \mathfrak{C}_x(s)}.$$

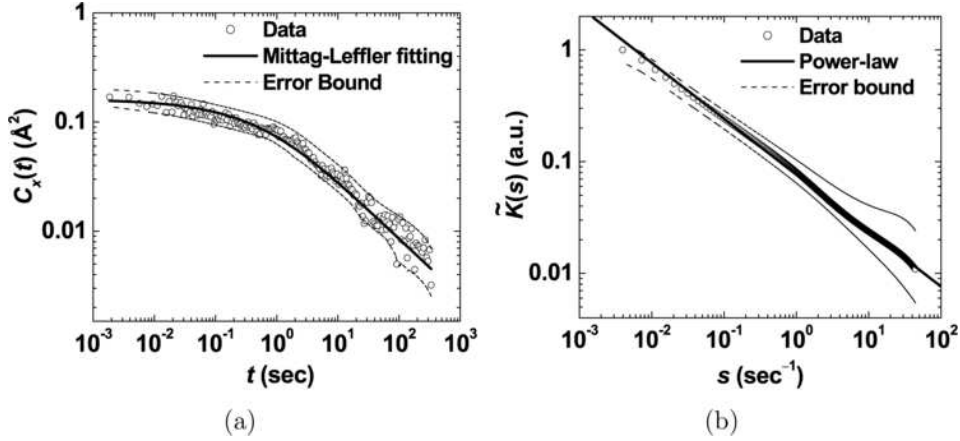

(a)                                            (b)

FIG. 7.   (a) *The autocorrelation functions of $x(t)$ calculated from the experimental data, the open circles, compared with the Mittag–Leffler expression (3.5) from our model, the solid line.* (b) *The experimentally determined memory kernel compared with the memory kernel of the fractional Gaussian noise.*

*Therefore, knowing* $C_x(t)$ *allows the recovery of* $K(t)$ *in the Laplace space.*

The proof of the theorem is given in the Appendix. With this theorem, using the empirical autocorrelation function $\tilde{C}_x(t)$, we can determine (in the Laplace space) the memory kernel experienced by the protein in the experiment. Figure 7(b) shows, in the Laplace space, the experimentally obtained memory kernel from (5.3) compared with the Laplace transform of the fractional Gaussian memory kernel $\mathfrak{K}_H(s) = \int_0^\infty e^{-st} K_H(t)\,dt = \Gamma(2H + 1)s^{1-2H}$, which is a power law. Close agreement is seen, which verifies the second key assumption of our model.

**6. Discussion.**   To explain the experimentally observed subdiffusion phenomenon, we formulate in this article a stochastic model by incorporating fractional Gaussian noise into the generalized Langevin equation framework. The resulting stochastic integro-differential equations driven by fractional Brownian motion are nonstandard. We study in detail these model equations. Using the analytical results, we show that the model leads to a satisfactory account for subdiffusion.

The model, in addition, has three attractive features: (1) The underlying theory is simple. First, compared with the classical Brownian diffusion theory, the model has only one more parameter: the Hurst parameter $H$. Second, the model offers analytical tractability. For instance, under the harmonic potential, closed form expressions of the displacement covariance function are obtained in Theorems 3.1 and 3.2. (2) The model, derivable from a Hamiltonian consideration, has a sound physical basis, which is an important requirement for biophysical models. (3) The theoretical results from the model agree well with the experimental data. Not only are the model predictions confirmed by the experiments, but also each key model assumption is directly verified in the experiments.

The successful application of our model to explain subdiffusion only exemplifies one instance of the numerous and growing stochastic modeling opportunities in nanoscale biophysics. Many interesting problems remain to be explored.

1. Existence, uniqueness and solution under a general potential. In this paper we solved the model equations (GLE with fGn) for the harmonic potential $U(x) = m\psi x^2/2$. For a general potential $U(x)$, the equation becomes

$$dx(t) = v(t)\,dt,$$

$$(6.1) \qquad m\,dv(t) = -\zeta \left( \int_{-\infty}^{t} v(u) K_H(t-u)\,du \right) dt$$

$$- U'(x(t))\,dt + \sqrt{2\zeta k_B T}\,dB_H(t).$$

A natural follow-up question is as follows: under what conditions does there exist such a bivariate process $(x(t), v(t))$, and when is it unique, for example, in the weak sense? Furthermore, if such a bivariate process exists and is unique, then how might one solve it, at least numerically? The answers to these questions are directly related to many biological and chemical systems, since many such biophysical and biochemical processes are subject to general potentials.

2. First passage time and rare events calculation. Many biological events are associated with the first passage time problem; for example, the completion of an enzymatic reaction corresponds to the first time that the enzymatic system escapes an energy barrier [Risken (1989), Van Kampen (2001)]. Consequently, calculating the distribution of the first passage time is of immediate biological applicability. For instance, the distribution of the first time that $x(t)$ reaches a given level from equation (6.1) directly relates to the understanding of enzymatic reactions in a sluggish protein environment [Min et al. (2005)]. Furthermore, these first passage time problems in biology usually correspond to the crossing of a very high barrier. An interesting open problem is thus to investigate how simulation, such as importance sampling, or large deviation techniques can be applied here to assess or approximate the probability of high barrier crossing.

3. Interacting particle systems. It is seen in Section 4 that our model can be derived from a system of interacting particles. This type of microscopic picture is not unusual for biological systems, as biological events, such as gene expression, tend to resolve from the interaction between many small units. It is thus interesting to see how the modern understanding of interacting particle systems can be extended to biological systems. For our model, in particular, a linear coupling between $q_j$ and $x$ in the Hamiltonian (4.2) is assumed in the derivation. It is desirable to relax this assumption to extend the microscopic derivation to more general interaction terms.

The booming field of nanoscale (single-molecule) biophysics has attracted much attention from biologists, chemists and physicists, as it projects a bright picture for new scientific discoveries. It also presents many interesting and challenging problems for stochastic modelers due to the stochastic nature of the nanometer world. It is our hope that this article will generate further interest in applying modern probabilistic and statistical methodology to interesting biophysical and scientific problems.

## APPENDIX: PROOFS AND DERIVATIONS

PROOF OF THEOREM 2.1. The pathwise interpretation of $dB_H(t)$ allows us to apply the technique of analyzing classical integro-differential equations to solve (2.9). Suppose $v(t)$ is the solution. Then applying a Fourier transform to both sides of equation (2.9), we know that $\tilde{v}(\omega) = \int_{-\infty}^{\infty} e^{it\omega} v(t)\, dt$ must satisfy

$$-im\omega \tilde{v}(\omega) = -\zeta \tilde{v}(\omega)\tilde{K}_H^+(\omega) + \sqrt{2\zeta k_B T} \int_{-\infty}^{\infty} e^{it\omega}\, dB_H(t).$$

The unique solution of the above equation is

$$\tilde{v}(\omega) = \sqrt{2\zeta k_B T} \int_{-\infty}^{\infty} e^{it\omega}\, dB_H(t) / [\zeta \tilde{K}_H^+(\omega) - im\omega].$$

An inverse Fourier transform gives

$$v(t) = \frac{1}{2\pi} \int_{-\infty}^{\infty} e^{-it\omega} \tilde{v}(\omega)\, d\omega = \sqrt{2\zeta k_B T} \int_{-\infty}^{\infty} r(t-u)\, dB_H(u).$$

Since $r(t)$ is deterministic, it is straightforward to verify that $v(t)$ given above is indeed a finite Gaussian process with zero mean. To calculate the covariance function, we use the fact [cf. Duncan, Hu and Pasik-Duncan (2000)] that, for $H > 1/2$ and deterministic functions $f$ and $g$,

$$E\left[ \int f(u)\, dB_H(u) \cdot \int g(u)\, dB_H(u) \right]$$
$$= \int\int H(2H-1)|u-v|^{2H-2} f(u)g(v)\, du\, dv,$$

which provides

$$E[v(t)v(s)]$$

(A.1)
$$= 2k_B T \zeta \int_{-\infty}^{\infty} \int_{-\infty}^{\infty} H(2H-1)|u-v|^{2H-2} r(t-u) r(s-v) \, du \, dv$$

$$= k_B T \zeta \int_{-\infty}^{\infty} \int_{-\infty}^{\infty} r(t-u) r(s-v) K_H(u-v) \, du \, dv.$$

A change of variable of $y = t - u$, $z = s - v$ gives

$$E[v(t)v(s)] = k_B T \zeta \int_{-\infty}^{\infty} \int_{-\infty}^{\infty} r(y) r(z) K_H(t-s-y+z) \, dy \, dz$$

$$= E[v(0)v(t-s)].$$

Therefore, $v(t)$ is a stationary Gaussian process. To obtain the stationary covariance $C_v(t) = E[v(0)v(t)]$, we compute $\tilde{C}_v(\omega) = \int_{-\infty}^{\infty} e^{it\omega} C_v(t) \, dt$.

From (A.1), it follows by a change of variable of $y = t - u$, $z = u - v$ that

$$\tilde{C}_v(\omega) = \int_{-\infty}^{\infty} e^{it\omega} C_v(t) \, dt$$

$$= k_B T \zeta \int_{-\infty}^{\infty} dt \, e^{it\omega} \int_{-\infty}^{\infty} \int_{-\infty}^{\infty} r(t-u) r(-v) K_H(u-v) \, du \, dv$$

$$= k_B T \zeta \left( \int_{-\infty}^{\infty} r(y) e^{iy\omega} \, dy \right) \left( \int_{-\infty}^{\infty} r(-v) e^{iv\omega} \, dv \right) \left( \int_{-\infty}^{\infty} K_H(z) e^{iz\omega} \, dz \right).$$

By the definition of $r(t)$, the above expression is simplified to

$$\tilde{C}_v(\omega) = k_B T \zeta \tilde{K}_H(\omega) / [(\zeta \tilde{K}_H^+(\omega) - i\omega m)(\zeta \tilde{K}_H^+(-\omega) + i\omega m)],$$

which for $\omega \in \mathsf{R}$ can be further simplified to

(A.2)
$$\tilde{C}_v(\omega) = k_B T \zeta \tilde{K}_H(\omega) / |\zeta \tilde{K}_H^+(\omega) - i\omega m|^2$$

because $\tilde{K}_H^+(-\omega)$ is the complex conjugate of $\tilde{K}_H^+(\omega)$ as the kernel function $K_H(t)$ is real. $\quad \square$

PROOF OF THEOREM 2.2. Since $E[x(t)] = \int_0^t E[v(s)] \, ds = 0$, it follows that, for $t > 0$,

$$\mathrm{Var}[x(t)] = E[x^2(t)] = \int_0^t \int_0^t E[v(s)v(u)] \, du \, ds = 2 \int_0^t \int_0^s C_v(u) \, du \, ds.$$

The right-hand side is

$$2 \int_0^t \int_0^s C_v(u) \, du \, ds = 2 \int_0^t \int_0^s \left[ \frac{1}{2\pi} \int_{-\infty}^{\infty} e^{-iu\omega} \tilde{C}_v(\omega) \, d\omega \right] du \, ds$$

$$= 2 \int_0^t \int_0^s \left[ \frac{1}{\pi} \int_0^{\infty} \cos(u\omega) \tilde{C}_v(\omega) \, d\omega \right] du \, ds.$$

Applying Fubini's theorem twice simplifies it to

$$2 \int_0^t \int_0^s C_v(u) \, du \, ds = \frac{2}{\pi} \int_0^\infty \frac{1}{\omega^2} \tilde{C}_v(\omega)(1 - \cos t\omega) \, d\omega.$$

Plugging in the result of Theorem 2.1, we obtain

$$E[x^2(t)] = \frac{2}{\pi} \int_0^\infty \frac{1}{\omega^2} \frac{k_B T \zeta \tilde{K}_H(\omega)}{|\zeta \tilde{K}_H^+(\omega) - im\omega^2|^2}(1 - \cos t\omega) \, d\omega$$

$$= \frac{2}{\pi} \int_0^\infty (2k_B T \zeta \Gamma(2H+1) \sin(H\pi) \omega^{-1-2H}(1 - \cos t\omega))$$
$$\times (m^2\omega^2 + 2\Gamma(2H+1)\cos(H\pi)m\zeta\omega^{2-2H}$$
$$+ \zeta^2\Gamma^2(2H+1)\omega^{2-4H})^{-1} \, d\omega.$$

A change of variable $\eta = t\omega$ gives

$$E[x^2(t)]$$
$$= \frac{2}{\pi} t^{2-2H} \int_0^\infty (2k_B T \zeta \Gamma(2H+1) \sin(H\pi) \eta^{-1-2H}(1 - \cos \eta))$$
$$\times (m^2\eta^2 t^{-4H} + 2\Gamma(2H+1)\cos(H\pi)m\zeta\eta^{2-2H}t^{-2H}$$
$$+ \zeta^2\Gamma^2(2H+1)\eta^{2-4H})^{-1} \, d\eta.$$

As $t \to \infty$, by the dominated convergence theorem, the above integral converges to

$$\int_0^\infty \frac{2k_B T \sin(H\pi)\eta^{-1-2H}(1 - \cos \eta)}{\zeta \Gamma(2H+1)\eta^{2-4H}} \, d\eta = \frac{k_B T}{\zeta} \frac{\sin(2H\pi)}{2H(1 - 2H)(2 - 2H)}.$$

Therefore, as $t \to \infty$, the mean-squared displacement

$$E[x^2(t)]/t^{2-2H} \to \frac{k_B T}{\zeta} \frac{\sin(2H\pi)}{\pi H(1 - 2H)(2 - 2H)}. \qquad \square$$

PROOF OF THEOREM 3.1. Applying a Fourier transform to equation (3.2), we know that $\tilde{x}(\omega) = \int_{-\infty}^\infty e^{it\omega} x(t) \, dt$ and $\tilde{v}(\omega) = \int_{-\infty}^\infty e^{it\omega} v(t) \, dt$ must satisfy

$$\tilde{v}(\omega) = -i\omega\tilde{x}(\omega),$$

$$-im\omega\tilde{v}(\omega) = -\zeta\tilde{v}(\omega)\tilde{K}_H^+(\omega) - m\psi\tilde{x}(\omega) + \sqrt{2\zeta k_B T} \int_{-\infty}^\infty e^{it\omega} \, dB_H(t),$$

which has the unique solution

$$\tilde{x}(\omega) = \frac{\sqrt{2\zeta k_B T}}{m\psi - m\omega^2 - i\omega\zeta\tilde{K}_H^+(\omega)} \int_{-\infty}^\infty e^{it\omega} \, dB_H(t),$$

$$\tilde{v}(\omega) = \frac{-\sqrt{2\zeta k_B T}i\omega}{m\psi - m\omega^2 - i\omega\zeta\tilde{K}_H^+(\omega)} \int_{-\infty}^\infty e^{it\omega} \, dB_H(t).$$

An inverse Fourier transform gives

$$x(t) = \sqrt{2\zeta k_B T} \int_{-\infty}^{\infty} \rho(t-u) \, dB_H(u),$$

$$v(t) = \sqrt{2\zeta k_B T} \int_{-\infty}^{\infty} \rho'(t-u) \, dB_H(u).$$

With $\rho(t)$ being deterministic, it is straightforward to verify that $(x(t), v(t))$ given above is a finite stationary bivariate Gaussian process with zero mean. For the covariance function $E[x(0)x(t)]$, we have

$$E[x(0)x(t)] = 2k_B T \zeta \int_{-\infty}^{\infty} \int_{-\infty}^{\infty} H(2H-1)|u-v|^{2H-2} \rho(t-u)\rho(-v) \, du \, dv$$

$$= k_B T \zeta \int_{-\infty}^{\infty} \int_{-\infty}^{\infty} \rho(t-u)\rho(-v) K_H(u-v) \, du \, dv.$$

The Fourier transform of the above equation is

$$\int_{-\infty}^{\infty} e^{it\omega} E[x(0)x(t)] \, dt$$

$$= k_B T \zeta \int_{-\infty}^{\infty} dt \, e^{it\omega} \int_{-\infty}^{\infty} \int_{-\infty}^{\infty} \rho(t-u)\rho(-v) K_H(u-v) \, du \, dv$$

$$= k_B T \zeta \left( \int_{-\infty}^{\infty} \rho(y) e^{iy\omega} \, dy \right) \left( \int_{-\infty}^{\infty} \rho(-v) e^{iv\omega} \, dv \right) \left( \int_{-\infty}^{\infty} K_H(z) e^{iz\omega} \, dz \right),$$

$$y = t - u, \qquad z = u - v.$$

By the definition of $\rho(t)$, the above expression is simplified to

$$\int_{-\infty}^{\infty} e^{it\omega} E[x(0)x(t)] \, dt = k_B T \zeta \tilde{K}_H(\omega)/|m\psi - m\omega^2 - i\omega\zeta \tilde{K}_H^+(\omega)|^2.$$

Thus,

$$E[x(0)x(t)] = \frac{k_B T \zeta}{2\pi} \int_{-\infty}^{\infty} e^{-it\omega} \tilde{K}_H(\omega)/|m\psi - m\omega^2 - i\omega\zeta \tilde{K}_H^+(\omega)|^2 \, d\omega.$$

The expressions of $E[x(0)v(t)]$, $E[v(0)x(t)]$ and $E[v(0)v(t)]$ can be obtained similarly.  □

PROOF OF THEOREM 3.2.   Following the proofs of Theorems 2.1 and 3.1, applying the Fourier method on equation (3.4) and some detailed calculations afterward yield

$$x(t) = \sqrt{2\zeta k_B T} \int_{-\infty}^{\infty} \mu(t-u) \, dB_H(u),$$

$$v(t) = \sqrt{2\zeta k_B T} \int_{-\infty}^{\infty} \mu'(t-u) \, dB_H(u),$$

where

$$\mu(t) = \frac{1}{2\pi} \int_{-\infty}^{\infty} e^{-it\omega} \frac{1}{m\psi - i\omega\zeta\tilde{K}_H^+(\omega)} \, d\omega,$$

from which we know that $E[x(t)] = 0$, and that the Fourier transform $\tilde{\sigma}_x(\omega) = \int_{-\infty}^{\infty} e^{it\omega} \sigma_x(t) \, dt = \int_{-\infty}^{\infty} e^{it\omega} E[x(0)x(t)] \, dt$ satisfies

$$\tilde{\sigma}_x(\omega) = k_B T \zeta \tilde{K}_H(\omega) / |m\psi - i\omega\zeta\tilde{K}_H^+(\omega)|^2.$$

Using the expressions (2.11) and (2.12), we have, for $\omega > 0$,

$$
\begin{aligned}
\text{(A.3)} \quad \tilde{\sigma}_x(\omega) &= (2k_B T \zeta \Gamma(2H+1)\sin(H\pi)\omega^{1-2H}) \\
&\quad \times (m^2\psi^2 - 2m\psi\zeta\Gamma(2H+1)\cos(H\pi)\omega^{2-2H} \\
&\qquad\qquad\qquad + \zeta^2\Gamma^2(2H+1)\omega^{4-4H})^{-1} \\
&= \frac{k_B T}{m\psi} \frac{2\sin(H\pi)(\tau\omega)^{2-2H}/\omega}{1 - 2\cos(H\pi)(\tau\omega)^{2-2H} + (\tau\omega)^{4-4H}},
\end{aligned}
$$

where the last equality uses the definition of $\tau$ in (3.6).

To establish (3.5), we only need to show that the Fourier transform of $\frac{k_B T}{m\psi} E_{2-2H}(-(t/\tau)^{2-2H})$ is exactly equal to $\tilde{\sigma}_x(\omega)$, that is, $\int_{-\infty}^{\infty} e^{-it\omega}(k_B T/m\psi) E_{2-2H}(-(t/\tau)^{2-2H}) \, dt = \tilde{\sigma}_x(\omega)$, which by (A.3) reduces to show that, for $\omega > 0$,

$$
\begin{aligned}
\text{(A.4)} \quad & 2\int_0^{\infty} \cos(t\omega) E_{2-2H}(-(t/\tau)^{2-2H}) \, dt \\
&= \frac{2\sin(H\pi)(\tau\omega)^{2-2H}/\omega}{1 - 2\cos(H\pi)(\tau\omega)^{2-2H} + (\tau\omega)^{4-4H}}.
\end{aligned}
$$

The Laplace transform of the Mittag–Leffler function has been given in Erdélyi et al. (1953), Chapter 18 as

$$\int_0^{\infty} e^{pt} E_\alpha(-(t/\tau)^\alpha) \, dt = \frac{1}{p} \frac{1}{1 + (\tau p)^{-\alpha}}.$$

Taking $p = i\omega$ in the above formula gives

$$
\begin{aligned}
& 2\int_0^{\infty} \cos(t\omega) E_{2-2H}(-(t/\tau)^{2-2H}) \, dt \\
&= 2\,\text{Re}\left(\frac{1}{i\omega} \frac{1}{1 + (i\tau\omega)^{-(2-2H)}}\right) \\
&= \frac{2}{\omega}\,\text{Re}\left(\frac{1}{i} \frac{1}{1 + (\tau\omega)^{-(2-2H)}e^{-i(1-H)\pi}}\right) \\
&= \frac{2}{\omega}\,\text{Re}\left(\frac{1}{i + (\tau\omega)^{-(2-2H)}(-i\cos(H\pi) + \sin(H\pi))}\right)
\end{aligned}
$$

$$= \frac{2}{\omega} \frac{\sin(H\pi)(\tau\omega)^{2-2H}}{1 - 2\cos(H\pi)(\tau\omega)^{2-2H} + (\tau\omega)^{4-4H}},$$

which is exactly (A.4). The proof is thus completed. $\square$

PROOF OF THEOREM 4.1.   From the result of Theorem 2.1, we have

$$\mathrm{Var}[v(0)] = E[v^2(0)]$$

$$= \frac{1}{2\pi} \int_{-\infty}^{\infty} k_B T \zeta \tilde{K}_H(\omega) / |\zeta \tilde{K}_H^+(\omega) - im\omega|^2 \, d\omega$$

$$= \frac{1}{\pi} \int_0^{\infty} (2k_B T \zeta \Gamma(2H+1)\sin(H\pi)\omega^{1-2H})$$

$$\times (\zeta^2 \Gamma^2(2H+1)\omega^{2-4H} + m^2\omega^2$$

$$+ 2m\zeta\Gamma(2H+1)\omega^{2-2H}\cos(H\pi))^{-1} \, d\omega.$$

A change of variable $\eta = \omega^{2H}$ gives

$$\mathrm{Var}[v(0)] = \frac{k_B T}{\pi H} \int_0^{\infty} \frac{\zeta\Gamma(2H+1)\sin(H\pi)}{m^2\eta^2 + 2m\zeta\Gamma(2H+1)\cos(H\pi)\eta + \zeta^2\Gamma^2(2H+1)} \, d\eta.$$

Using the general formula $\int_0^{\infty} \frac{dx}{x^2 + 2xy\cos\phi + y^2} = \frac{\phi}{y\sin\phi}$, the above expression is simplified to

$$\mathrm{Var}[v(0)] = \frac{k_B T}{m} \qquad \text{for all } 1/2 < H < 1,$$

which agrees with the thermal dynamic requirement. $\square$

PROOF OF THEOREM 4.2.   Theorem 3.1 implies

$$\mathrm{Var}[x(0)]$$

$$= E[x^2(0)]$$

$$= \frac{k_B T \zeta}{2\pi} \int_{-\infty}^{\infty} \tilde{K}_H(\omega) / |m\psi - m\omega^2 - i\omega\zeta\tilde{K}_H^+(\omega)|^2 \, d\omega$$

$$= \frac{k_B T}{\pi} \int_0^{\infty} (2\zeta\Gamma(2H+1)\sin(H\pi)\omega^{1-2H})$$

$$\times ([m\psi - m\omega^2 - \zeta\Gamma(2H+1)\omega^{2-2H}\cos(H\pi)]^2$$

$$+ [\zeta\Gamma(2H+1)\omega^{2-2H}\sin(H\pi)]^2)^{-1} \, d\omega.$$

A change of variable $\eta = \tau\omega$, where $\tau = [\zeta\Gamma(2H+1)]^{1/(2-2H)}$, gives

$$\mathrm{Var}[x(0)] = \frac{k_B T}{\pi} \int_0^{\infty} (2\sin(H\pi)\eta^{1-2H})$$

(A.5) $\qquad\qquad \times ([m\psi - m\eta^2/\tau^2 - \eta^{2-2H}\cos(H\pi)]^2$
$$+ [\eta^{2-2H}\sin(H\pi)]^2)^{-1}\,d\eta.$$

Consider the complex valued function

$$f(z) = [z(m\psi - mz^2/\tau^2 - z^{2-2H}e^{iH\pi})]^{-1}.$$

It is straightforward to verify that it is analytic on the region defined by the boundary curve

$$C = [1/R, R] \cup \{Re^{i\theta} : 0 \le \theta \le \pi\} \cup [-R, -1/R] \cup \{e^{i\theta}/R : 0 \le \theta \le \pi\},$$

where the real number $R > 1$, and $[1/R, R]$ is the real interval between $1/R$ and $R$.

It follows that $\int_C f(z)\,dz = 0$. But

$$\int_C f(z)\,dz = I + II + III + IV,$$

where

$$I = \int_{1/R}^R [x(m\psi - mx^2/\tau^2 - x^{2-2H}e^{iH\pi})]^{-1}\,dx,$$

$$II = \int_0^\pi [Re^{i\theta}(m\psi - mR^2 e^{2i\theta}/\tau^2 - R^{2-2H}e^{i\theta(2-2H)}e^{iH\pi})]^{-1}\,dRe^{i\theta},$$

$$III = \int_{-R}^{-1/R} [x(m\psi - mx^2/\tau^2 - x^{2-2H}e^{iH\pi})]^{-1}\,dx,$$

$$IV = \int_\pi^0 \left[\frac{e^{i\theta}}{R}\left(m\psi - m\frac{e^{2i\theta}}{R^2\tau^2} - \frac{e^{i\theta(2-2H)}}{R^{2-2H}}e^{iH\pi}\right)\right]^{-1}\,d\frac{e^{i\theta}}{R}.$$

We thus have $I + III = -(II + IV)$. We can simplify $I + III$, $II$ and $IV$ as

$$I + III = \int_{1/R}^R [x(m\psi - mx^2/\tau^2 - x^{2-2H}e^{iH\pi})]^{-1}\,dx$$

$$+ \int_{1/R}^R [x(m\psi - mx^2/\tau^2 - x^{2-2H}e^{-iH\pi})]^{-1}\,dx$$

$$= \int_{1/R}^R \frac{1}{x}\frac{2i\sin(H\pi)x^{2-2H}}{[m\psi - mx^2/\tau^2 - x^{2-2H}\cos(H\pi)]^2 + [x^{2-2H}\sin(H\pi)]^2}\,dx,$$

$$II = i\int_0^\pi (m\psi - mR^2 e^{2i\theta}/\tau^2 - R^{2-2H}e^{i\theta(2-2H)}e^{iH\pi})^{-1}\,d\theta,$$

$$IV = -i\int_0^\pi \left(m\psi - m\frac{e^{2i\theta}}{R^2\tau^2} - \frac{e^{i\theta(2-2H)}}{R^{2-2H}}e^{iH\pi}\right)^{-1}\,d\theta.$$

It therefore follows that

$$\int_{1/R}^{R} \frac{1}{x} \frac{2\sin(H\pi)x^{2-2H}}{[m\psi - mx^2/\tau^2 - x^{2-2H}\cos(H\pi)]^2 + [x^{2-2H}\sin(H\pi)]^2}\,dx$$

$$= \int_0^\pi \left(m\psi - m\frac{e^{2i\theta}}{R^2\tau^2} - \frac{e^{i\theta(2-2H)}}{R^{2-2H}}e^{iH\pi}\right)^{-1}d\theta$$

$$- \int_0^\pi (m\psi - mR^2 e^{2i\theta}/\tau^2 - R^{2-2H}e^{i\theta(2-2H)}e^{iH\pi})^{-1}\,d\theta.$$

Letting $R \to +\infty$ provides (by the dominated convergence theorem)

$$\lim_{R\to+\infty}\int_0^\pi \left(m\psi - m\frac{e^{2i\theta}}{R^2\tau^2} - \frac{e^{i\theta(2-2H)}}{R^{2-2H}}e^{iH\pi}\right)^{-1}d\theta = \frac{\pi}{m\psi},$$

$$\lim_{R\to+\infty}\int_0^\pi (m\psi - mR^2 e^{2i\theta}/\tau^2 - R^{2-2H}e^{i\theta(2-2H)}e^{iH\pi})^{-1}\,d\theta = 0,$$

yielding

$$\int_0^\infty \frac{1}{x}\frac{2\sin(H\pi)x^{2-2H}}{[m\psi - mx^2/\tau^2 - x^{2-2H}\cos(H\pi)]^2 + [x^{2-2H}\sin(H\pi)]^2}\,dx = \frac{\pi}{m\psi}.$$

Plugging this expression into (A.5), we finally obtain

$$E[x^2(0)] = \frac{k_B T}{m\psi}\qquad\text{for all } 1/2 < H < 1.\qquad\qquad\square$$

PROOF OF THEOREM 5.1. Consider the function

$$\breve{h}(s) = \frac{k_B T \zeta}{m\psi}\frac{\mathfrak{K}(s)}{m\psi + \zeta s \mathfrak{K}(s)},$$

where $\mathfrak{K}(s)$ is the Laplace transform of the memory kernel $K(t)$. The inverse Laplace transform $h(t)$ of $\breve{h}(s)$ is given by Doetsch (1974), pages 4 and 148,

$$h(t) = \frac{1}{2\pi i}\int_{-i\infty}^{i\infty}e^{st}\breve{h}(s)\,ds$$

$$= \frac{1}{2\pi}\int_{-\infty}^{\infty}e^{i\omega t}\breve{h}(i\omega)\,d\omega,\qquad \omega = is$$

$$= \frac{2}{\pi}\int_0^\infty \cos(\omega t)\,\mathrm{Re}[\breve{h}(i\omega)]\,d\omega.$$

Since

$$\mathrm{Re}[\breve{h}(i\omega)] = \frac{k_B T \zeta}{m\psi}\,\mathrm{Re}\left[\frac{\mathfrak{K}(i\omega)}{m\psi + i\zeta\omega\mathfrak{K}(i\omega)}\right] = \frac{k_B T\zeta\,\mathrm{Re}[\mathfrak{K}(i\omega)]}{|m\psi - i\zeta\omega\mathfrak{K}(-i\omega)|^2}$$

and

$$\tilde{K}(\omega) = \int_{-\infty}^{\infty} e^{it\omega} K(t) \, dt = 2 \operatorname{Re}[\mathfrak{K}(i\omega)],$$

$$\tilde{K}^{+}(\omega) = \int_{0}^{\infty} e^{it\omega} K(t) \, dt = \mathfrak{K}(-i\omega),$$

we obtain

$$h(t) = \frac{1}{\pi} \int_{0}^{\infty} \cos(\omega t) k_B T \zeta \tilde{K}(\omega) / |m\psi - i\zeta\omega \tilde{K}^{+}(\omega)|^2 \, d\omega.$$

On the other hand, from the proof of Theorem 3.2, we know that for a general memory kernel $K(t)$, the covariance $C_x(t) = \operatorname{cov}(x(0), x(t))$ is given by

$$C_x(t) = \frac{1}{2\pi} \int_{-\infty}^{\infty} e^{-it\omega} k_B T \zeta \tilde{K}(\omega) / |m\psi - i\omega\zeta \tilde{K}^{+}(\omega)|^2 \, d\omega$$

$$= \frac{1}{\pi} \int_{0}^{\infty} \cos(\omega t) k_B T \zeta \tilde{K}(\omega) / |m\psi - i\zeta\omega \tilde{K}^{+}(\omega)|^2 \, d\omega,$$

which is identical to $h(t)$. Therefore, the Laplace transform $\mathfrak{C}_x(s)$ of $C_x(t)$ is $\breve{h}(s)$, namely,

$$\mathfrak{C}_x(s) = \frac{k_B T \zeta}{m\psi} \frac{\mathfrak{K}(s)}{m\psi + \zeta s \mathfrak{K}(s)}. \qquad \square$$

*Exact expressions of the higher-order autocorrelation functions of* $\lambda(t)$. Since the fluorescence lifetime $\lambda(t)$ and the distance fluctuation $x(t)$ are linked by

$$\lambda(t) = k_0 e^{\beta(x_{eq} + x(t))},$$

to calculate the higher-order autocorrelations of $\lambda(t)$, the following expression is very useful.

*A useful expression.* Suppose $x(t)$ is a stationary Gaussian process with mean 0, and covariance function $C_x(t) = \operatorname{Cov}(x(0), x(t))$. Then for any $t_1 \leq t_2 \leq \cdots \leq t_n$, the expectation of $E\{e^{Ax(t_1)} e^{Ax(t_2)} \cdots e^{Ax(t_n)}\}$, where $A$ is a constant, is given by

$$(A.6) \quad E\{e^{Ax(t_1)} e^{Ax(t_2)} \cdots e^{Ax(t_n)}\} = \exp\left\{\frac{n}{2} A^2 C_x(0) + A^2 \sum_{i<j} C_x(t_j - t_i)\right\}.$$

PROOF. Since $x(t_1) + \cdots + x(t_n)$ is Gaussian with mean 0 and variance $n C_x(0) + 2 \sum_{i<j} C_x(t_j - t_i)$, it follows that $e^{Ax(t_1)} e^{Ax(t_2)} \cdots e^{Ax(t_n)}$ is lognormally distributed, and the standard result of the log-normal distribution yields

$$E\{e^{Ax(t_1)} e^{Ax(t_2)} \cdots e^{Ax(t_n)}\} = \exp\left\{\frac{n}{2} A^2 C_x(0) + A^2 \sum_{i<j} C_x(t_j - t_i)\right\}.$$

$\square$

Using this expression, we first have $E[\lambda(t)] \equiv E[\lambda(0)] = E[k_0 e^{\beta(x_{eq}+x(t))}] = k_0 \exp(\beta x_{eq} + \frac{1}{2}\beta^2 C_x(0))$. For the three-step correlation function, the stationarity of $\lambda(t)$ reduces $E[\Delta\lambda(0)\Delta\lambda(t_1)\Delta\lambda(t_1+t_2)]$ to

$$E[\Delta\lambda(0)\Delta\lambda(t_1)\Delta\lambda(t_1+t_2)]$$
$$= E[\{\lambda(0) - E[\lambda(0)]\}\{\lambda(t_1) - E[\lambda(t_1)]\}\{\lambda(t_1+t_2) - E[\lambda(t_1+t_2)]\}]$$
$$= E\{\lambda(0)\lambda(t_1)\lambda(t_1+t_2)\} + 2\{E[\lambda(0)]\}^3$$
$$- E[\lambda(0)](E\{\lambda(0)\lambda(t_1)\} + E\{\lambda(0)\lambda(t_1+t_2)\} + E\{\lambda(t_1)\lambda(t_1+t_2)\}).$$

Using (A.6), it is further simplified to

$$E[\Delta\lambda(0)\Delta\lambda(t_1)\Delta\lambda(t_1+t_2)]$$
$$= k_0^3 e^{3\beta x_{eq}+3\beta^2 C_x(0)/2}\{e^{\beta^2[C_x(t_1)+C_x(t_2)+C_x(t_1+t_2)]} - e^{\beta^2 C_x(t_1)}$$
$$- e^{\beta^2 C_x(t_2)} - e^{\beta^2 C_x(t_1+t_2)} + 2\}.$$

Similarly, expanding the individual terms in the four-step correlation function and using the stationarity and (A.6) provide

$$E[\Delta\lambda(0)\Delta\lambda(t_1)\Delta\lambda(t_1+t_2)\Delta\lambda(t_1+t_2+t_3)]$$
$$= k_0^4 e^{4\beta x_{eq}+2\beta^2 C_x(0)}$$
$$\times \{e^{\beta^2[C_x(t_1)+C_x(t_2)+C_x(t_3)+C_x(t_1+t_2)+C_x(t_2+t_3)+C_x(t_1+t_2+t_3)]}$$
$$- e^{\beta^2[C_x(t_1)+C_x(t_2)+C_x(t_1+t_2)]}$$
$$- e^{\beta^2[C_x(t_1)+C_x(t_2+t_3)+C_x(t_1+t_2+t_3)]}$$
$$- e^{\beta^2[C_x(t_1+t_2)+C_x(t_3)+C_x(t_1+t_2+t_3)]}$$
$$- e^{\beta^2[C_x(t_2)+C_x(t_3)+C_x(t_2+t_3)]} + e^{\beta^2 C_x(t_1)}$$
$$+ e^{\beta^2 C_x(t_2)} + e^{\beta^2 C_x(t_3)} + e^{\beta^2 C_x(t_1+t_2)}$$
$$+ e^{\beta^2 C_x(t_2+t_3)} + e^{\beta^2 C_x(t_1+t_2+t_3)} - 3\}.$$

**Acknowledgments.** The author thanks the Xie group at the Department of Chemistry and Chemical Biology of Harvard University for fruitful discussions. The author is also grateful to Professors Chris Heyde, Ruth Williams and Robert Adler for helpful comments.

## REFERENCES

ADLER, R., FELDMAN, R. and TAQQU, M. (1998). *A Practical Guide to Heavytails*: *Statistical Techniques for Analyzing Heavy-Tailed Distributions*. Birkhäuser, Boston.

ALÒS, E., MAZET, O. and NUALART, D. (2000). Stochastic calculus with respect to fractional Brownian motion with Hurst parameter less than 1/2. *Stochastic Process. Appl.* **86** 121–139. MR1741199

ASBURY, C., FEHR, A. and BLOCK, S. M. (2003). Kinesin moves by an asymmetric hand-over-hand mechanism. *Science* **302** 2130–2134.

BOUCHAUD, J. and GEORGES, A. (1990). Anomalous diffusion in disordered media: Statistical mechanisms, models and physical applications. *Phys. Rep.* **195** 127–293. MR1081295

CARMONA, P. and COUTIN, L. (2000). Intégrale stochastique pour le mouvement brownien fractionnaire. *C. R. Acad. Sci. Paris* **330** 231–236. MR1748314

CHAMPENEY, D. C. (1987). *A Handbook of Fourier Theorems*. Cambridge Univ. Press. MR0900583

CHANDLER, D. (1987). *Introduction to Modern Statistical Mechanics*. Oxford Univ. Press, New York. MR0913936

CORBEN, H. C. and STEHLE, P. (1995). *Classical Mechanics*. Dover Publications, New York. MR1298889

CROVELLA, M. and BESTAVROS, A. (1996). Self-similarity in world wide web traffic: Evidence and possible causes. *Performance Evaluation Review* **24** 160–169.

DAI, W. and HEYDE, C. C. (1996). Ito's formula with respect to fractional Brownian motion and its application. *J. Appl. Math. Stochast. Anal.* **9** 439–448. MR1429266

DOETSCH, G. (1974). *Introduction to the Theory and Application of the Laplace Transformation*. Springer, New York. MR0344810

DUNCAN, T. E., HU, Y. and PASIK-DUNCAN, B. (2000). Stochastic calculus for fractional Brownian motion I. Theory. *SIAM J. Control Optim.* **38** 582–612. MR1741154

EMBRECHTS, P. and MAEJIMA, M. (2002). *Selfsimilar Processes*. Princeton Univ. Press. MR1920153

ENGLISH, B., MIN, W., VAN OIJEN, A. M., LEE, K. T., LUO, G., SUN, H., CHERAYIL, B. J., KOU, S. C. and XIE, X. S. (2006). Ever-fluctuating single enzyme molecules: Michaelis–Menten equation revisited. *Nature Chemical Biology* **2** 87–94.

ERDÉLYI, A. et al. (1953). *High Transcendental Functions* **3**. McGraw-Hill, New York.

GLYNN, P. and ZEEVI, A. (2000). On the maximum workload of a queue fed by fractional Brownian motion. *Ann. Appl. Probab.* **10** 1084–1099. MR1810865

GRAY, H. and WINKLER, J. (1996). Electron transfer in proteins. *Annu. Rev. Biochem.* **65** 537–561.

GRIPENBERG, G. and NORROS, I. (1996). On the prediction of fractional Brownian motion. *J. Appl. Probab.* **33** 400–410. MR1385349

HEATH, D., RESNICK, S. and SAMORODNITSKY, G. (1997). Patterns of buffer overflow in a class of queues with long memory in the input stream. *Ann. Appl. Probab.* **7** 1021–1057. MR1484796

HEYDE, C. C. (1999). A risky asset model with strong dependence through fractal activity time. *J. Appl. Probab.* **36** 1234–1239. MR1746407

HILL, T. (1986). *An Introduction to Statistical Thermodynamics*. Dover, New York. MR0951633

KARLIN, S. and TAYLOR, H. (1981). *A Second Course in Stochastic Processes*. Academic Press, New York. MR0611513

KLAFTER, J., SHLESINGER, M. and ZUMOFEN, G. (1996). Beyond Brownian motion. *Physics Today* **49** 33–39.

KONSTANTOPOULOS, T. and LIN, S. J. (1996). Fractional Brownian approximations of queueing networks. *Stochastic Networks. Lecture Notes in Statist.* **117** 257–273. Springer, New York. MR1466791

KOU, S. C., CHERAYIL, B., MIN, W., ENGLISH, B. and XIE, X. S. (2005). Single-molecule Michaelis-Menten equations. *J. Phys. Chem. B* **109** 19068–19081.

KOU, S. C. and XIE, X. S. (2004). Generalized Langevin equation with fractional Gaussian noise: Subdiffusion within a single protein molecule. *Phys. Rev. Lett.* **93** 180603(1)–180603(4).

KOU, S. C., XIE, X. S. and LIU, J. S. (2005). Bayesian analysis of single-molecule experimental data (with discussion). *J. Roy. Statist. Soc. Ser. C* **54** 469–506. MR2137252

KOU, S. C. (2007). Stochastic networks in nanoscale biophysics: Modeling enzymatic reaction of a single protein. *J. Amer. Statist. Assoc.* To appear.

KUPFERMAN, R. (2004). Fractional kinetics in Kac–Zwanzig heat bath models. *J. Statist. Phys.* **114** 291–326. MR2032133

LELAND, W. E., TAQQU, M. S., WILLINGER, W. and WILSON, D. V. (1994). On the self-similar nature of Ethernet traffic (Extended Version). *IEEE/ACM Trans. Networking* **2** 1–15.

LIN, S. J. (1995). Stochastic analysis of fractional Brownian motions. *Stochast. Stochast. Rep.* **55** 121–140. MR1382288

LU, H. P., XUN, L. and XIE, X. S. (1998). Single-molecule enzymatic dynamics. *Science* **282** 1877–1882.

MANDELBROT, B. (1997). *Fractals and Scaling in Finance.* Springer, New York. MR1475217

MANDELBROT, B. and VAN NESS, J. (1968). Fractional Brownian motions, fractional noises and applications. *SIAM Rev.* **10** 422–437. MR0242239

MIKOSCH, T. and NORVAISA, R. (2000). Stochastic integral equations without probability. *Bernoulli* **6** 401–434. MR1762553

MIKOSCH, T., RESNICK, S., ROOTZÉN, H. and STEGEMAN, A. (2002). Is network traffic approximated by stable Lévy motion or fractional Brownian motion? *Ann. Appl. Probab.* **12** 23–68. MR1890056

MIN, W., ENGLISH, B., LUO, G., CHERAYIL, B., KOU, S. C. and XIE, X. S. (2005). Fluctuating enzymes: Lessons from single-molecule studies. *Accounts of Chemical Research* **38** 923–931.

MIN, W., LUO, G., CHERAYIL, B., KOU, S. C. and XIE, X. S. (2005). Observation of a power law memory kernel for fluctuations within a single protein molecule. *Phys. Rev. Lett.* **94** 198302(1)–198302(4).

MOERNER, W. (2002). A dozen years of single-molecule spectroscopy in physics, chemistry, and biophysics. *J. Phys. Chem. B* **106** 910–927.

MOSER, C., KESKE, J., WARNCKE, K., FARID, R. and DUTTON, P. (1992). Nature of biological electron transfer. *Nature* **355** 796–802.

MUKAMEL, S. (1995). *Principle of Nonlinear Optical Spectroscopy.* Oxford Univ. Press, New York.

NIE, S. and ZARE, R. (1997). Optical detection of single molecules. *Ann. Rev. Biophys. Biomol. Struct.* **26** 567–596.

NUALART, D. (2006). *The Malliavin Calculus and Related Topics* (*Probability and Its Applications*). Springer, New York. MR2200233

PIPIRAS, V. and TAQQU, M. S. (2000). Integration questions related to fractional Brownian motion. *Probab. Theory Related Fields* **118** 251–291. MR1790083

PIPIRAS, V. and TAQQU, M. S. (2001). Are classes of deterministic integrands for fractional Brownian motion on an interval complete? *Bernoulli* **7** 873–897. MR1873833

Reif, F. (1965). *Fundamentals of Statistical and Thermal Physics*. McGraw-Hill, New York.

Risken, H. (1989). *The Fokker–Planck Equation*: *Methods of Solution and Applications*. Springer, Berlin. MR0987631

Rogers, L. C. G. (1997). Arbitrage with fractional Brownian motion. *Math. Finance* **7** 95–105. MR1434408

Samorodnitsky, G. and Taqqu, M. (1994). *Stable Non-Gaussian Random Processes*. Chapman and Hall, New York. MR1280932

Shiryaev, A. N. (1998). On arbitrage and replication for fractal models. Research Report No. 2, 1998, MaPhySto, Univ. Aarhus.

Sokolov, I., Klafter, J. and Blumen, A. (2002). Fractional kinetics. *Physics Today* **55** 48–54.

Tamarat, P., Maali, A., Lounis, B. and Orrit, M. (2000). Ten years of single-molecule spectroscopy. *J. Phys. Chem. A* **104** 1–16.

Taqqu, M. S. (1986). Sojourn in an elliptical domain. *Stochastic Process. Appl.* **21** 319–326. MR0833958

Van Kampen, N. G. (2001). *Stochastic Processes in Physics and Chemistry*. North-Holland, Amsterdam.

Wang, K. G. and Tokuyama, M. (1999). Nonequilibrium statistical description of anomalous diffusion. *Phys. A* **265** 341–351.

Weiss, S. (2000). Measuring conformational dynamics of biomolecules by single molecule fluorescence spectroscopy. *Nature Struct. Biol.* **7** 724–729.

Whitt, W. (2002). *Stochastic-Process Limits*. Springer, New York. MR1876437

Xie, X. S. and Lu, H. P. (1999). Single-molecule enzymology. *J. Bio. Chem.* **274** 15967–15970.

Xie, X. S. and Trautman, J. K. (1998). Optical studies of single molecules at room temperature. *Ann. Rev. Phys. Chem.* **49** 441–480.

Yang, H., Luo, G., Karnchanaphanurach, P., Louise, T.-M., Rech, I., Cova, S., Xun, L. and Xie, X. S. (2003). Protein conformational dynamics probed by single-molecule electron transfer. *Science* **302** 262–266.

Zhuang, X., Kim, H., Pereira, M., Babcock, H., Walter, N. and Chu, S. (2002). Correlating structural dynamics and function in single ribozyme molecules. *Science* **296** 1473–1476.

Zwanzig, R. (2001). *Nonequilibrium Statistical Mechanics*. Oxford Univ. Press, New York. MR2012558

Department of Statistics
Harvard University
1 Oxford Street
Cambridge, Massachusetts 02138
USA
E-mail: kou@stat.harvard.edu